\documentclass[11pt,preprint]{aastex}
\usepackage{graphicx}
\usepackage{amsmath,amssymb} 
\usepackage{subfigure}

\begin{document}

\newcommand{\rhat}{\hat{\boldsymbol r}}
\newcommand{\that}{\hat{\boldsymbol \theta}}
\newcommand{\zhat}{\hat{\boldsymbol z}}
\newcommand{\phat}{\hat{\boldsymbol \phi}}
\newcommand{\bfnabla}{\boldsymbol \nabla}
\newcommand{\bfv}{\boldsymbol v}
\newcommand{\bfomega}{\boldsymbol \omega}
\newcommand{\bfB}{\boldsymbol B}
\newcommand{\bfT}{\boldsymbol T}
\newcommand{\ba}{\begin{eqnarray}}
\newcommand{\ea}{\end{eqnarray}}
\newcommand{\eh}{\hat{\boldsymbol e}}
\newcommand{\cp}{\varpi}
\newcommand{\ol}{\overline}
\newcommand{\sgn}{\mathop{\rm sgn}}
\newcommand{\ephi}{{\bf e}_\varphi}
\newcommand{\Fdis}{F_{\rm w,dis}}
\newcommand{\eps}{\epsilon}
\newcommand{\La}{\langle}
\newcommand{\Ra}{\rangle}


\title{Angular Momentum Transport by Acoustic Modes
  Generated in the Boundary Layer II: MHD Simulations.}

\author{Mikhail A. Belyaev\altaffilmark{1} \& Roman
  R. Rafikov\altaffilmark{1} \& James M. Stone\altaffilmark{1}}
\altaffiltext{1}{Department of Astrophysical Sciences, 
Princeton University, Ivy Lane, Princeton, NJ 08540; 
rrr@astro.princeton.edu}


\begin{abstract}
We perform global unstratified 3D magnetohydrodynamic simulations of
an astrophysical boundary layer (BL) -- an interface region between an
accretion disk and a weakly magnetized accreting object such as a 
white dwarf -- with the goal of understanding the effects of magnetic
field on the BL. We use cylindrical 
coordinates with an isothermal equation of state and investigate a 
number of initial field geometries including toroidal, vertical, and 
vertical with zero net flux. Our initial setup consists of a
Keplerian disk attached to a non-rotating star. In a previous work, we
found that in hydrodynamical simulations, sound waves excited by shear in the
BL were able to efficiently transport angular momentum and drive mass
accretion onto the star. Here we confirm that in MHD simulations,
waves serve as an efficient means of angular momentum transport in 
the vicinity of the BL, despite the magnetorotational instability 
(MRI) operating in the disk. In particular, the angular momentum
current due to waves is at times larger than the angular momentum current
due to MRI. Our results suggest that angular momentum transport in the BL 
and its vicinity is a global phenomenon occurring through
dissipation of waves and shocks. This point of view is quite different from the
standard picture of transport by a local anomalous turbulent viscosity. In
addition to angular momentum transport, we also study
magnetic field amplification within the BL. We find that the field is
indeed amplified in the BL, but only by a factor of a few and remains
subthermal.
\end{abstract}



\section{Introduction}

The mechanism of angular momentum transport in the boundary layers
(BLs) of accreting
compact objects and stars is an open problem that is both interesting
from a physical point of view and important for modeling
BL spectra from these sources \citep{Narayan, PophamNarayan1, PophamNarayan,
  InogamovSunyaev, PiroBildsten}. In this work, we shall focus on the
case of an accretion disk that does not undergo disruption by the 
magnetic field of an accreting opbject
(for the case of magnetically disrupted disk see \citet{GhoshLamb}, 
  \citet{Koldobaetal}) and extends all the way to the surface of the
compact object or star. Such a situation is expected to be naturally 
realized in weakly magnetized neutron stars producing Type-I X-ray 
bursts as well as during FU Orioni outbursts in pre-main sequence stars 
and dwarf nova outbursts in cataclysmic variables when the mass 
accretion rate is high.

Accretion of matter through the BL must be accompanied by the 
angular momentum transport in the layer, but the nature of
this transport has remained elusive. The BL is linearly stable to the
magnetorotational instability
(MRI; \cite{Velikhov,Chandra,BalbusMRI}) that is thought to
give rise to anomalous viscosity in the bulk of well-ionized 
accretion disks, because the angular
velocity, $\Omega$, necessarily {\it increases} with radius in
this region. Of course linear stability does not imply a nonturbulent
flow. Turbulence,
once generated by some nonlinear mechanism, can persist
without decaying in certain classes of rotating, linearly
stable hydrodynamical flows \citep{LesurLongaretti}.
However transport mechanisms, both magnetic and hydrodynamic
involving an anomalous viscosity have never
been definitively demonstrated to operate inside the BL.

Recently \citet{BR} and \citet{BRS,BRS1} showed both theoretically and
computationally that
non-axisymmetric acoustic modes are excited in the BL when it is thin
in comparison with the stellar radius. These
acoustic modes manifest themselves as sound waves that
propagate away from the boundary layer, efficiently transporting
angular momentum into both the star and the disk and driving
accretion. However, the studies of \citet{BR}, \citet{BRS,BRS1} were
purely hydrodynamical, and it is the purpose of
the present study to check their results on angular momentum
transport by waves generated in the BL in the presence of
a magnetic field.

There are several reasons why
the MHD case could be fundamentally different from the hydrodynamical
case. First, the presence of a seed magnetic field in the disk leads to
turbulence generated by the MRI. Potentially, the turbulence could 
interact with the acoustic modes in the disk, modifying their
properties or washing them out completely. Second, magnetic field advected
into the BL by accreting gas will be amplified by the shear present there
\citep{Pringle1989, Armitage}. A strongly magnetized BL could have 
different properties than an unmagnetized one, again influencing the 
properties of acoustic modes. Third, magnetic field in the BL can give 
rise to other MHD effects such as the Tayler-Spruit dynamo
\citep{Tayler,Spruit1999,Spruit2002}, which may produce additional 
momentum transport in the BL \citep{PiroBildsten07}.

One of our aims in this work is to determine the extent to which a
magnetic field affects the properties of and angular momentum
transport by waves generated in the BL. We
find that in the presence of a magnetic field, the acoustic modes
studied by \citet{BRS,BRS1} become magnetosonic modes. However, when
the ratio of gas pressure to magnetic pressure $\beta \gg 1$, the
magnetic field
has only a weak, $\mathcal{O}\left(\beta^{-1}\right)$ effect on the
dynamics of acoustic modes. In our simulations, $\beta$ is indeed
large, and we find that magnetosonic modes play an important role
in angular momentum transport in the BL and its vicinity. 

Previously, \citet{Armitage} and \citet{SteinackerPapaloizou} each
performed MHD simulations of the BL. However, these studies did not 
reveal the importance of the transport by waves, and we 
discuss our work in relation to theirs in \S \ref{discussion}.
Also, \citet{HP}, \citet{HP1}, \citet{HP2} studied angular momentum
transport by spiral density waves in MRI turbulent disks,
which were sourced by vorticity generated in non-linear
MRI turbulence. However, the mechanism for driving these waves is
entirely different
from the mechanism for generating the magnetosonic modes discussed
here. The latter are sourced by
shear in the BL and launched away from it into both the star and the
disk.

There have also been a number of other analytical and numerical studies of
disk wave phenomena in the vicinity of the innermost stable circular orbit 
of a black hole and near the BL \citep{LGN, LN, LaiTsang2009, TsangLai0, 
TsangLai, TsangLai1, FuLai2012}. However, these studies were primarily 
addressing the possible connection between the non-axisymmetric modes
and quasi-periodic oscillations observed in accreting objects, whereas
our main focus is on the fundamental nature of the angular momentum 
transport within the BL. Moreover, with the exception of 
\citet{TsangLai} and \citet{LN}, these studies did not include 
a magnetic field.

The paper is organized as follows. In \S \ref{nummod}, we discuss the
technical aspects of our numerical model, and in \S
\ref{gensec} we introduce notation and define quantities used throughout
the paper when discussing angular momentum transport. Then, in \S
\ref{disksec} we talk about angular momentum transport by MRI in the
disk and show that our results agree with previous studies of the
MRI. \S \ref{acoussec} discusses
how acoustic modes studied by \citet{BRS,BRS1} are modified by the
presence of a magnetic field, becoming magnetosonic modes. Finally, \S
\ref{angmomwave} discusses
angular momentum transport by these magnetosonic modes in the star, BL,
and inner disk.

\section{Numerical Model}
\label{nummod}
We begin by summarizing the equations and the numerical model used in
our simulations.
In this paper, we use a numerical model that is identical to the one
presented in \citet{BRS1}, except for the addition of a magnetic
field. Thus, we will only briefly summarize the aspects of the
numerical model that do not relate to the magnetic field, referring
the reader to \citet{BRS1} for rest of the details.

\subsection{Governing Equations and Nondimensional Quantities}
For our simulations, we use Athena \citep{Stone} to solve
the MHD equations with an isothermal
equation of state in cylindrical geometry. These equations are
\ba
\label{eq:cont}
\frac{\partial \rho}{\partial t} + \bfnabla \cdot (\rho \bfv) &=& 0 \\
\label{eq:mom}
\frac{\partial (\rho \bfv)}{\partial t} + \bfnabla \cdot (\rho \bfv
\bfv) &=& -\bfnabla \left( P + \frac{B^2}{2\mu} \right) + \frac{1}{\mu}
(\bfB \cdot \bfnabla) \bfB - \rho \bfnabla \Phi \\
\label{eq:induction}
\frac{\partial \bfB}{\partial t} &=& \bfnabla \times (\bfv \times
\bfB) \\
\label{eq:state}
P &=& \rho s^2.
\ea
Here, $\bfv$ is the velocity, $\rho$ is the density, $P$ is the
pressure, $\Phi$ is the
gravitational potential, $s$ is the sound speed, and $\bfB$ is the
magnetic field. We also use $\cp$ throughout this work to denote the
cylindrical radius.

Adoption of an isothermal equation of state is equivalent 
to assuming  fast cooling to a set temperature. Although this is not a
realistic assumption, it allows us to simply attain
a quasi-steady state and focus on exploring the effect of a magnetic
field on the dynamics of the BL.

We nondimensionalize quantities by setting the radius of the star
to $\cp_\star = 1$, the Keplerian angular velocity in the disk just
above the surface of the star to $\Omega_K(\cp_\star) = 1$, the
density in the disk, initially a constant, to $\rho = 1$, and
the magnetic permeability to $\mu = 1$. Finally, we define a
characteristic Mach number for our
simulations as $M = 1/s$. This corresponds to the true Mach number of
fluid rotating at the Keplerian velocity at $\cp = \cp_\star$.

\subsection{Summary of Aspects of the Numerical Model not Relating to
  the Magnetic Field}

The simulation setup consists of a radially stratified, unrotating
star attached to a Keplerian disk via a thin interfacial region. The
simulations are performed in cylindrical geometry with an
isothermal equation of state and are not vertically
stratified. The interfacial region smoothly connects the rotation
profile of the star to that of the disk and is chosen to be as thin as
possible, while still being resolved with $\sim 10$ cells. The highest
density in the star is approximately $4 \times 10^6$ the density in
the disk. Thus, the inner regions of the star have a high enough
inertia to avoid being spun up over the course of a simulation.

We do not incorporate self-gravity into our runs, but instead use a
fixed gravitational potential $V = -1/\cp$. We initialize each of
our simulations to be in hydrostatic equilibrium and add random
perturbations to the radial velocity from which instabilities
develop. For additional information on aspects of our numerical model
that do not pertain to the magnetic field see \citet{BRS1}.

\subsection{Initial Magnetic Field Configuration}
\label{fieldconfsec}

We now discuss the initial magnetic field strengths and geometries used in
our simulations. For all our simulations, we set $\bfB = 0$
for $\cp < \cp_\star$ (an unmagnetized star), and initialize one of three
possible magnetic field geometries for $\cp \ge \cp_\star$:
\ba
\label{Bfieldinit}
\bfB(\cp \ge \cp_\star,z) = \left\{
     \begin{array}{lc}
       \frac{B_0}{\cp} \zhat, &  \text{NVF} \\
       B_0 \phat, & \text{NAF}\\
       \frac{B_0}{\cp} \sin\left[\frac{2 \pi}{\lambda_\cp} (\cp -
       \cp_\star)\right] \zhat, & \text{ZNF}
     \end{array}
   \right. .
\ea 
NVF simulation have a constant vertical magnetic flux
through any radius, NAF simulations have a constant azimuthal magnetic
flux, and ZNF simulations have on average zero
net flux. For the ZNF simulation, we set $\lambda_\cp = \Delta z/2$,
where $\Delta z$ is the extent of the simulation domain in the
vertical direction. Thus, we have a rapid variation of the field in
the radial direction, which allows comparison with shearing box ZNF
simulations.

In all of our runs, we use an initial value of $B_0 = .002$ in units for
which the magnetic permeability is $\mu = 1$. Defining the plasma
parameter $\beta$ as
\ba
\label{betadef}
\beta \equiv \frac{\rho s^2}{B^2/2\mu}
\ea
the initial value of $\beta$ using $B_0$ for a baseline is $\beta_0
\approx 6200$. Note that in the NVF and
ZNF simulations, $\beta$ varies as a function of position and
$\beta_0$ specifies the minimum initial value of $\beta$ in the disk.  

The particular value of $B_0 = .002$ is motivated by two
considerations. The first is the need
to resolve the wavelength of the fastest growing mode of the MRI in the
case of a pure vertical field (NVF simulations), and the second
is the need to fit several wavelengths of the fastest growing mode
within the simulation domain. 

For a global disk simulation in which
the domain has constant width and resolution in the $z$ direction,
these two criteria are generally at odds with one another. This is
because the $z$-wavelength of the fastest growing axisymmetric mode
for constant
vertical field in a Keplerian disk is given by \citet{BalbusMRI}
\ba
\label{MRIlambda}
\lambda_{z,\text{MRI}} = \sqrt{\frac{16}{15}} \frac{2 \pi B}{\Omega
  \sqrt{\rho}}.
\ea
For a Keplerian disk with constant $B_z$ field and constant $\rho$,
  $\lambda_{z,\text{MRI}} \propto \cp^{3/2}$. For our NVF field
  geometry, the magnetic field strength falls off
as $1/\cp$ so $\lambda_{z,\text{MRI}} \propto \cp^{1/2}$. This is a weaker
dependence than for a constant field and is the main reason we use a
constant flux criterion in the NVF case, rather than simply using a
constant field. 

For the ZNF case, we no longer have to worry about fitting the vertical
wavelength of the fastest growing mode within the simulation domain,
since the field strength varies as a function of radius and the MRI
grows on all scales. The same goes for NAF simulations,
because instability manifests
itself differently if the initial field is azimuthal rather than
vertical. In the former case, the most unstable modes are
non-axisymmetric ones that undergo transient exponential growth in the
shearing sheet approximation \citep{Balbus92,TPC} or pure exponential
growth in a global disk analysis
\citep{MatsumotoTajima,CKK,OgilviePringle,TerquemPapaloizou}. In a
shearing sheet analysis, the modes which undergo the most vigorous
growth are the ones that have the largest values of $k_z$ for a given
value of $k_x$ and $k_y$ \citep{Balbus92,HGB95}. Thus, resolution of
  these modes in a simulation is limited by the grid resolution, but their
  vertical wavelength is always able to fit within the vertical extent
  of the simulation domain.

\subsection{Simulation Parameters}
\label{simpar}

The only physical parameter varied across simulations in this study is
the initial field
geometry. In principle, other parameters such as the Mach number, the
vertical extent of the simulation domain,
the presence of vertical stratification, etc., would also have an
important effect on the dynamics of the BL.
However, it is computationally expensive to undertake an exhaustive
parameter study,
so we limit ourselves to varying only the initial field geometry in this
exploratory work. This allows us to explore the major effects a
magnetic field has on BL dynamics with no added frills. We mention
that \citet{BRS,BRS1} have already explored how varying the
dimensionality (2D/3D), stratification (stratified/unstratified), and
Mach number affects the outcome of hydrodynamic BL simulations.

Some important parameters that are
fixed across different runs are the Mach number, set to $M = 9$; the
vertical extent of the simulation domain, set to $\Delta z =
s/\Omega_K(\cp_\star) = 1/9$; the $\phi$ extent of the simulation domain,
set to $ \Delta \phi  = 2 \pi / 7$; and the $\phi$ resolution
set to $N_\phi = 384$ cells. Table
\ref{MHDsimtable} lists important
numerical parameters that vary from simulation to simulation.  

The specific choice of $ \Delta \phi  = 2
\pi / 7$, represents a compromise between computational resources and
simulating a wedge that spans a wide enough azimuthal angle. Also,
setting $ \Delta \phi  = 2 \pi / 7$, we resolve all modes having $m$ a
multiple of seven. The $m=14$ mode happens to be close to a geometric
resonance (equations (44)-(45) of
\citet{BRS1}) for $M=9$, which means its amplitude is enhanced
relative to modes
with a similar $m$ number. Therefore, the $m=14$ mode typically peaks
in our simulations, simplifying the analysis, since we can study a
single dominant mode rather than a superposition of modes having
similar amplitudes. Note that the finite extent of our simulations
in the azimuthal direction precludes us from exploring
the Tayler-Spruit dynamo \citep{Tayler,Spruit1999,Spruit2002} 
in the BL, which would require long azimuthal wavelength to be
resolved. However, the very existence in the BL of this dynamo 
previously shown to operate only in the incompressible limit 
should be seriously questioned in a highly dynamic, compressible 
environment such as the BL.

The boundary conditions we use are periodic in the $z$ and
$\phi$ directions, and do-nothing at the inner radial boundary. A
do-nothing boundary
condition means that the fluid quantities retain their initial values
on the boundary for the duration of the simulation. This is
preferred over an open (outflow) boundary condition for the inner
boundary, since
it ensures that the star does not slowly ``drift'' out of the
simulation domain. It is also preferred to a reflecting boundary
condition, since it is (partially) transparent to outgoing waves
\citep{BRS1}.

At the outer radial boundary, we implement two different types of
boundary conditions. The first type, which we call OBC0, is simply an open
(outflow) boundary condition. The second type, OBC1, is a do-nothing
boundary condition, just like at the inner radial boundary. In
simulations implementing an outer do-nothing boundary condition, we
initialize a nonmagnetic ``buffer zone'' between a fiducial radius in
the disk $\cp_\text{B,max}$ and the outer boundary. The magnetic field
is initialized to be zero in the region $\cp_\text{B,max} \le \cp \le
\cp_\text{out}$, where $\cp_\text{out}$ denotes the radius of the
outer boundary. In all of our simulations implementing OBC1,
$\cp_\text{B,max} = 4$.

\begin{figure}[!h]
\centering
\subfigure[]{\includegraphics[width=0.8\textwidth]{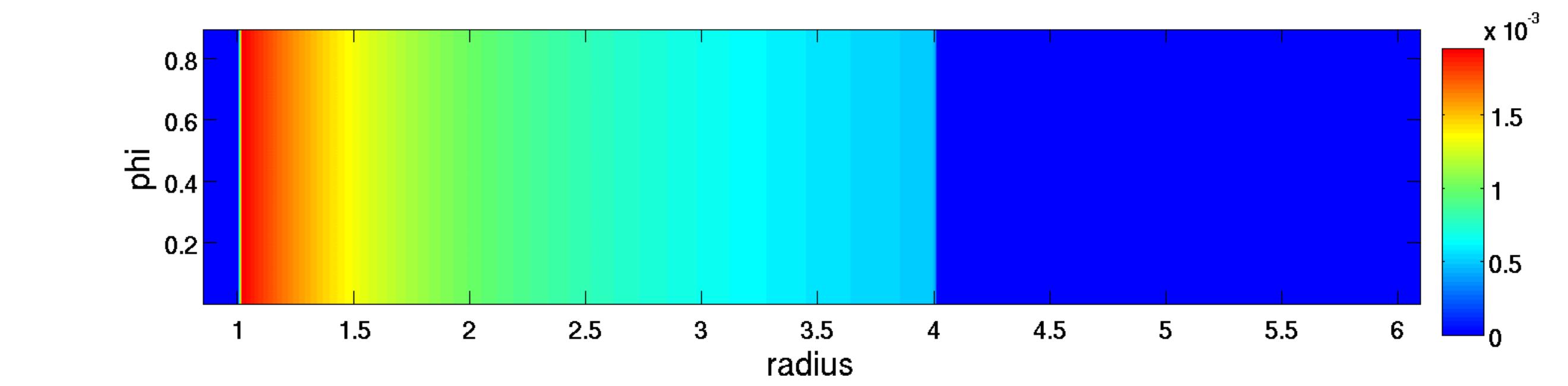}}
\subfigure[]{\includegraphics[width=0.8\textwidth]{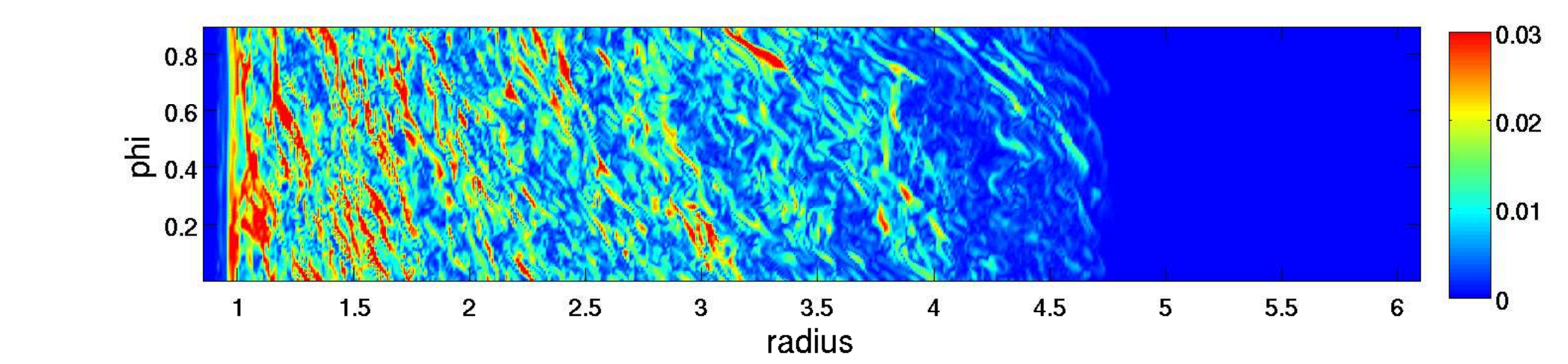}}
\caption{Panels a,b show the magnitude of the magnetic field in
  simulation M9d at times $t = 0$ and $t = 500$, respectively.}
\label{bufferfig}
\end{figure}

To demonstrate the buffer zone more clearly, we plot the magnetic
field magnitude from simulation M9d at $t = 0$ and $t = 500$ in panels
a and b of Fig. \ref{bufferfig}. In panel a, the field decays $\propto 1/\cp$
between the star at $\cp = 1$ and the buffer zone at $\cp =
\cp_\text{B,max} = 4$. In the region $\cp_\text{B,max} < \cp <
\cp_\text{out}$, the field is initially set to zero. In panel b, magnetic field
has penetrated into the buffer zone, as a result of the natural evolution of
the MHD equations, but has not yet reached the outer boundary. In general,
simulations implementing OBC1 are stopped before the magnetic field can diffuse
through the buffer zone and reach the outer boundary. This allows
us to run simulations for a long time, without having to worry about the
effect of magnetic field at the outer boundary on the rest of the
simulation domain.

\begin{table}[!h]
\centering
\begin{tabular}{|c|c|c|c|c|c|c|}
\hline
label & {\bf B} type & outer BC & $\cp$-range
& $N_\cp$ & $N_z$ & $t_\text{run}$ \\
\hline
M9a & NVF & OBC0 & (.85, 6.1)  & 4096 & 128 & 450 \\
M9b & NAF & OBC0 & (.85, 7.85) & 6144 & 64 & 450 \\
M9c & ZNF & OBC0 & (.85, 6.1) & 4096 & 128 & 450 \\
M9d & NVF & OBC1 & (.85, 6.1) & 4096 & 64 & 800 \\
M9e & NAF & OBC1 & (.85, 6.1) & 4096 & 48 & 1000 \\
M9f & ZNF & OBC1 & (.85, 6.1) & 4096 & 64 & 1800 \\
\hline
\end{tabular}
\caption{Summary of simulation parameters for MHD simulations. The
  columns from left to
  right are: simulation label, initial field geometry
  (net vertical flux, net azimuthal flux, or zero net flux), outer
  boundary condition, radial extent of the simulation domain,
  number of grid points in the radial direction, number of grid points
  in the vertical direction, and the total time for which the
  simulation was run.}
\label{MHDsimtable}
\end{table}  

\section{Angular Momentum Transport: Equations}
\label{gensec}

In this section, we introduce the notation used throughout this paper
when discussing angular momentum transport. When
analyzing simulation results, we
perform both density weighted and volume weighted averages of
quantities over the $z$ and $\phi$ directions. A volume weighted
average is denoted by an overline:
\ba
\overline{f} \equiv \frac{1}{\Delta z \Delta \phi}\int dz d\phi f, 
\ea
where $\Delta z$ and $\Delta \phi$ are the extents of the simulation
domain in the $z$ and $\phi$ dimensions respectively.
A density weighted average is denoted by angle brackets:
\ba
\La f \Ra \equiv \frac{1}{\Delta z \Delta \phi \Sigma_0}\int dz d\phi
\rho f, 
\ea
where $\Sigma_0 = \overline{\rho}$ is the volume weighted average
density. 

For ideal MHD, the one dimensional equation of angular momentum
transport can be formulated in terms of density weighted and volume
weighted averages as \citep{BalbusPapaloizou}
\ba
\label{1Dangmom}
\frac{\partial}{\partial t}\left(\cp^2 \Sigma_0 \Omega \right) +
\frac{1}{\cp} \frac{\partial}{\partial \cp} \left(\cp^3 \Omega \Sigma_0 \La
v_\cp \Ra + \cp^2  \overline{T_{\cp\phi}}
\right) &=& 0,
\ea
where $\Omega \equiv \La v_\phi \Ra/\cp$ is the density-weighted
angular velocity, and $T_{\cp\phi}$ is the
$\cp$, $\phi$ component of the stress tensor
\ba
T_{\cp\phi} \equiv \rho v_\cp \left(v_\phi - \cp \Omega\right) -
\frac{B_\cp B_\phi}{\mu}.
\ea

The first term in the space derivative in equation (\ref{1Dangmom})
equals the advected angular momentum current, $C_A$, divided by $2\pi$,
and the second term equals the
stress angular momentum current, $C_S$, divided by $2 \pi$.
For a steady state disk, $C_A + C_S = 0$, although the disks in our
simulations are not in steady state, and undergo temporal evolution.

The stress current, $C_S$, can be further decomposed into
hydrodynamical and magnetic terms
\ba
\label{CSH}
C_{S,H}(\cp,t) &\equiv& 2 \pi \cp^2 \Sigma_0 \La v_\cp \left(v_\phi -
\cp \Omega\right) \Ra \\
\label{CSB}
C_{S,B}(\cp,t) &\equiv& -2 \pi \cp^2 \overline{\frac{B_\cp B_\phi}{\mu}} \\
\label{CSsplit}
C_S &=& C_{S,H} + C_{S,B}.
\ea
Note that $C_S$ includes contributions to the stress from both waves
and the action of turbulence.

Finally, we define an $\alpha$ parameter to parametrize the
angular momentum transport rate by turbulent viscosity. There are many possible
definitions for $\alpha$, especially in the context of the BL
\citep{Narayan, PophamNarayan}, but the ones used in this work are
\ba
\label{alphaeq}
\alpha(\cp,t) &\equiv& \frac{\overline{T_{\cp\phi}}}{\overline{P}} =
\frac{\overline{T_{\cp\phi}}}{\Sigma_0 s^2} \\
\label{alphaSSeq}
\alpha_\nu(\cp,t) &\equiv& \frac{\overline{T_{\cp\phi}}}{\Sigma_0
  s H d\Omega/d \ln \cp}.
\ea
The first form of $\alpha$ is just the ratio of the stress to the
pressure. The second is the $\alpha$ for a turbulent
anomalous viscosity \citep{ShakuraSunyaev, LyndenBellPringle}, where
$H$ is the vertical scale of the problem.
In an astrophysical disk, $H  = s/\Omega$
and increases radially for constant $s$. Our simulations,
however, are in cylindrical geometry and
$H = \Delta z$, where $\Delta z$, the extent of the simulation
domain in the vertical direction, is a constant.

Analogous to the splitting of $C_S$ into hydrodynamical and magnetic
terms (equation [\ref{CSsplit}]), it is useful to split $\alpha$ into
hydrodynamical and magnetic components:
\ba
\alpha_H(\cp,t) &\equiv& \frac{\La v_\cp \left(v_\phi - \cp
    \Omega \right) \Ra}{s^2} \\
\alpha_B(\cp,t) &\equiv& -\frac{\overline{B_\cp B_\phi}}{\mu \Sigma_0 s^2} \\
\alpha &=& \alpha_H + \alpha_B.
\ea

Although $\alpha$ is a useful parametrization of angular momentum
transport by MRI
turbulence in the disk, it is not applicable to transport by
waves. In the latter case, the relevant quantity is $C_S$
\citep{BRS1}, and since
angular momentum is transported by both waves and MRI turbulence in
our simulations, we need to consider both $C_S$ and $\alpha$ when
analyzing results.

Another useful parameter we refer to often when analyzing simulations
is $\beta$ (equation [\ref{betadef}]), and from now on we take
\ba
\beta(\cp,t) \equiv \frac{\Sigma_0 s^2}{\overline{B^2}/2\mu}.
\ea

\section{Angular Momentum Transport by MRI in the Disk}
\label{disksec}

In this section, we focus on angular momentum transport in the disk and
show that we obtain resolved MRI
turbulence in our simulations. Thus, this section
serves as a check of our simulations against previous work, and the
reader who is primarily interested in new results about wave transport
of angular momentum may choose to skip ahead to the next section.

\subsection{Magnetic Pitch Angle}
\label{alphabetasec}

The primary diagnostic we use to check that the disk reaches a
resolved quasi-steady state of MRI turbulence is the magnetic pitch angle
\ba
\label{pitchangleeq}
\theta_B \equiv \frac{\sin^{-1}\left(\alpha_B \beta\right)}{2}.
\ea
Equation (\ref{pitchangleeq}) gives an approximation for the angle of the
magnetic field with respect to the $\phi$ direction if $B_\phi \gg
B_\cp$ and $B_\cp \gg B_z$
\citep{GuanGammie}. \citet{Sorathia} found that $\theta_B$ was a
particularly good indicator of convergence, since it varies monotonically with
resolution and converges to a value of $\theta_B \approx 13^{\circ}$
in resolved simulations. A similar indicator of convergence is the
relation $\alpha \beta \approx 1/2$ \citep{GuanGammie,HGK,Sorathia}. However,
$\alpha$ includes both hydrodynamical and magnetic stresses, whereas
$\alpha_B$ is calculated using only magnetic stresses. We find in our
simulations that the hydrodynamical stress due to waves is
non-negligible in the disk, but waves do not contribute to the
magnetic stress. Thus, a diagnostic based on $\alpha_B$ is strongly
preferred to one based on $\alpha$, when assessing the convergence of
MRI turbulence in a simulation that includes a BL.

\begin{figure}[!h]
\centering
\subfigure{\includegraphics[width=0.8\textwidth]{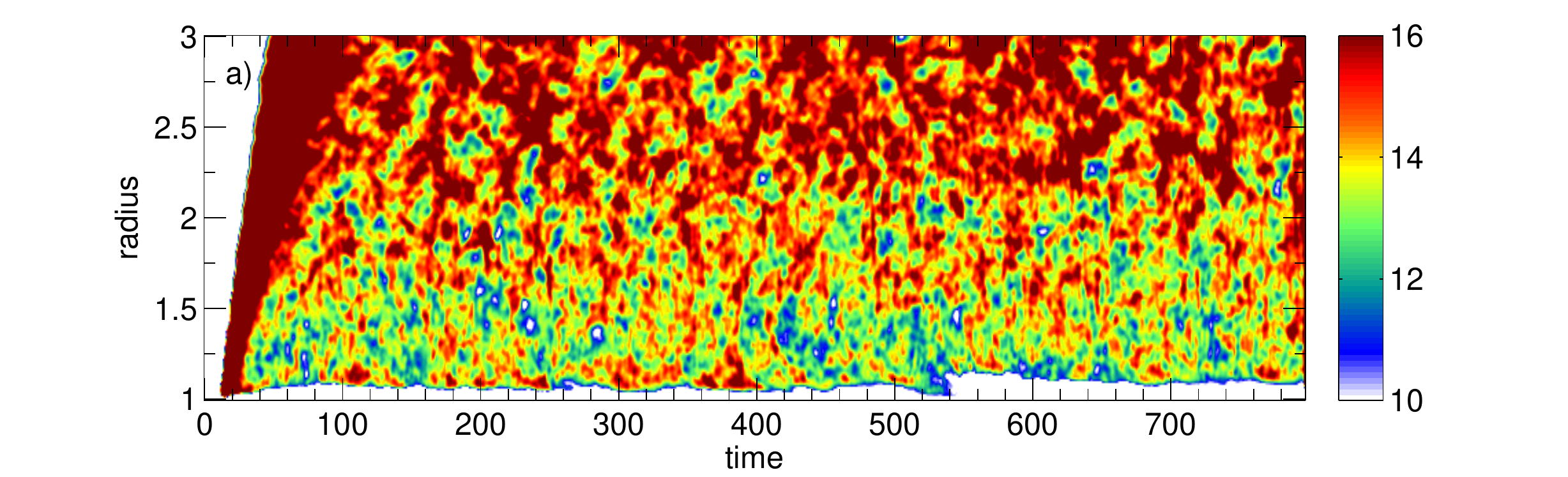}}
\subfigure{\includegraphics[width=0.8\textwidth]{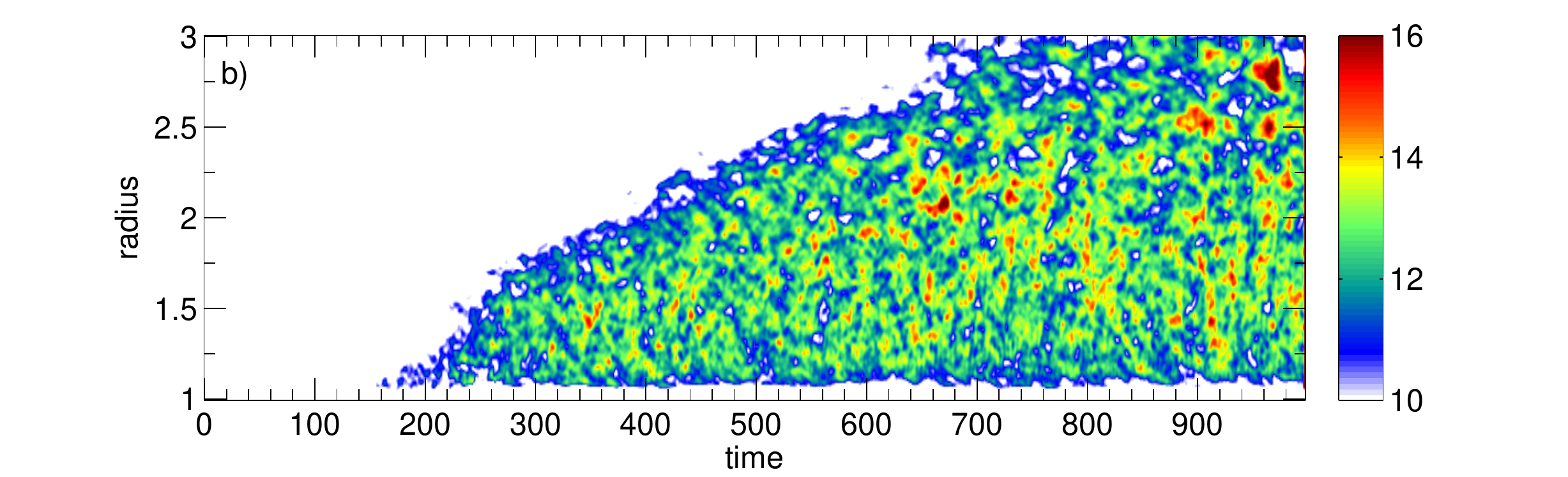}}
\subfigure{\includegraphics[width=0.8\textwidth]{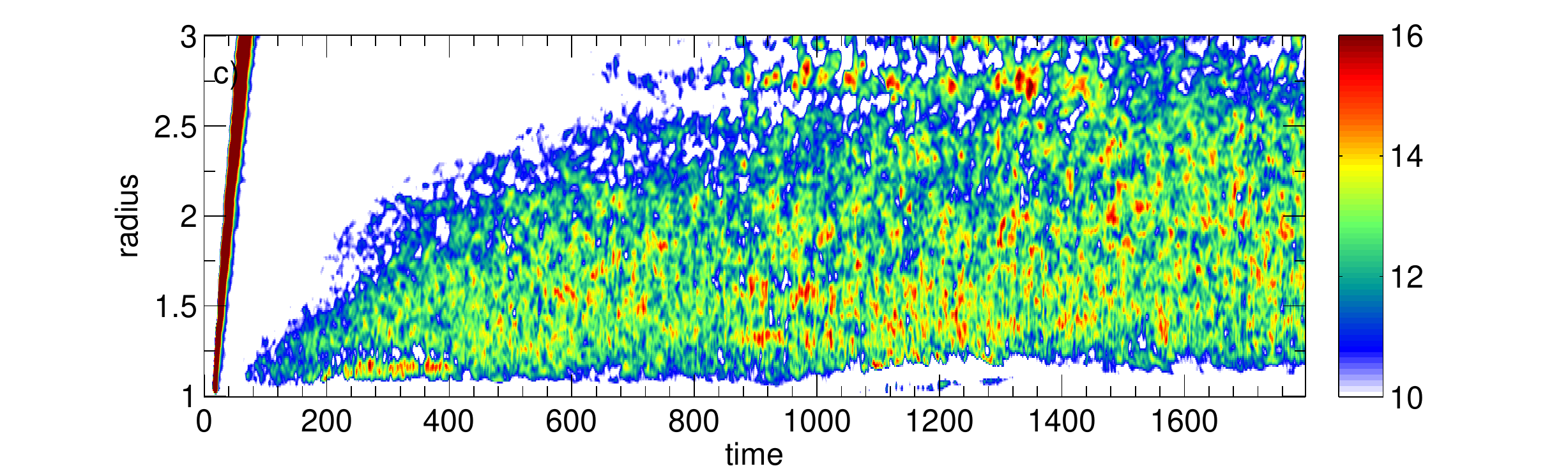}}
\caption{Panels a,b,c show $\theta_B$ for simulations M9d (NVF), M9e
  (NAF), and M9f (ZNF).  The band on the left hand side of panel c is due
  to a transient burst of activity that occurs when the MRI first goes
  nonlinear. Note the different scale on the time axis in
  the figures.}
\label{pitchanglespacetime}
\end{figure}

Fig. \ref{pitchanglespacetime} shows radius time plots of $\theta_B$
for simulations M9d,e,f in the annulus $1 \le \cp \le 3$. Ignoring
the region near $\cp = \cp_\star = 1$ where the rotation
profile is strongly non-Keplerian, we see
that $\theta_B \approx 12-14^\circ$ in the disk proper for the NAF and ZNF
simulations (panels b and c) once MRI turbulence is fully developed. In
the NVF simulations, $\theta_B \approx 12-14^\circ$ in the inner disk, but
rises with radius. Nevertheless, the values of $\theta_B$ observed in
our simulations are in good agreement with \citet{Sorathia}. This
gives us confidence that we have properly resolved the MRI in the
disk.

\subsection{Convergence Across Different Resolutions}
\label{convergesec}

\begin{figure}[!h]
\centering
\subfigure[]{\includegraphics[width=0.49\textwidth]{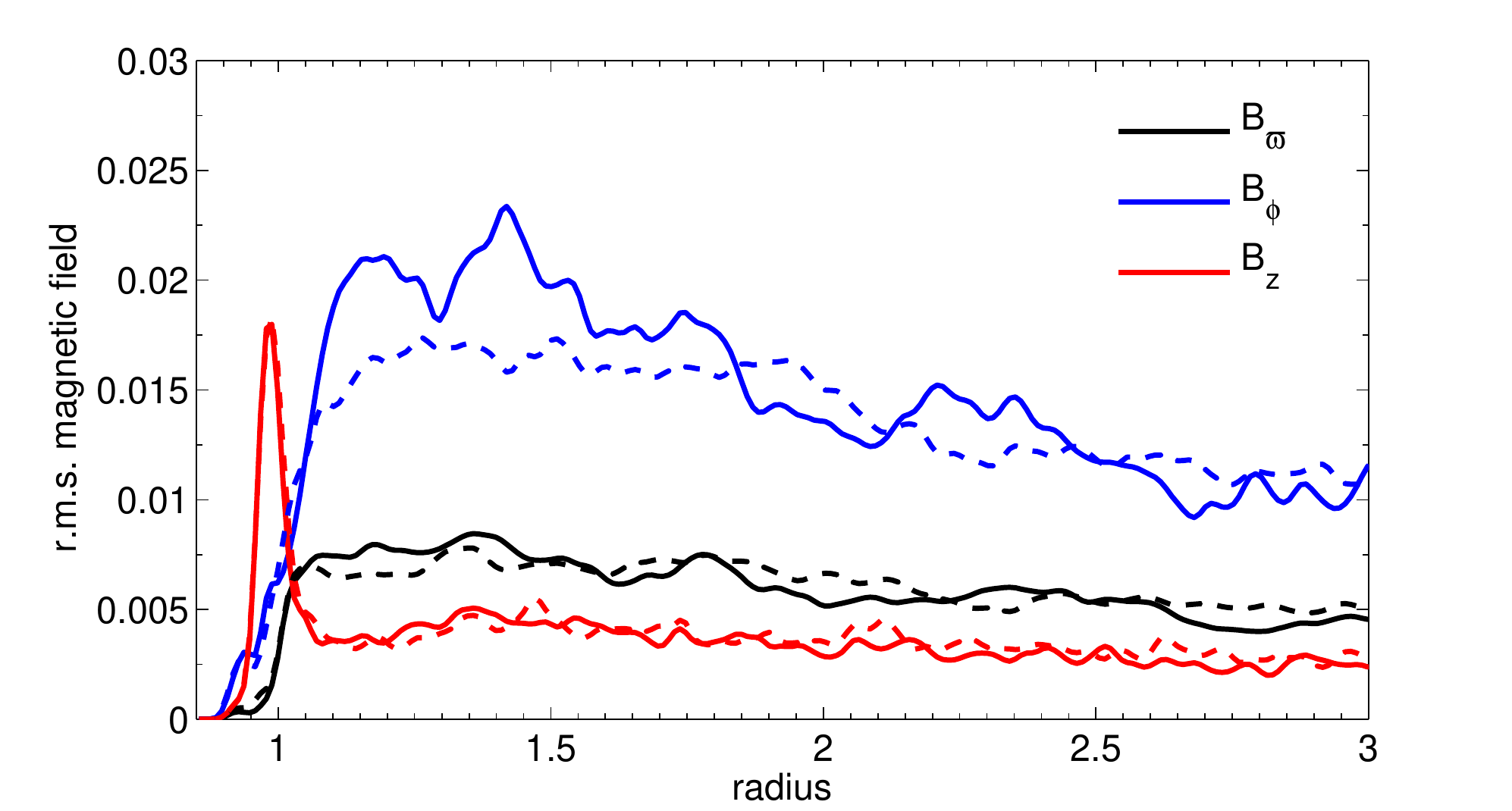}}
\subfigure[]{\includegraphics[width=0.49\textwidth]{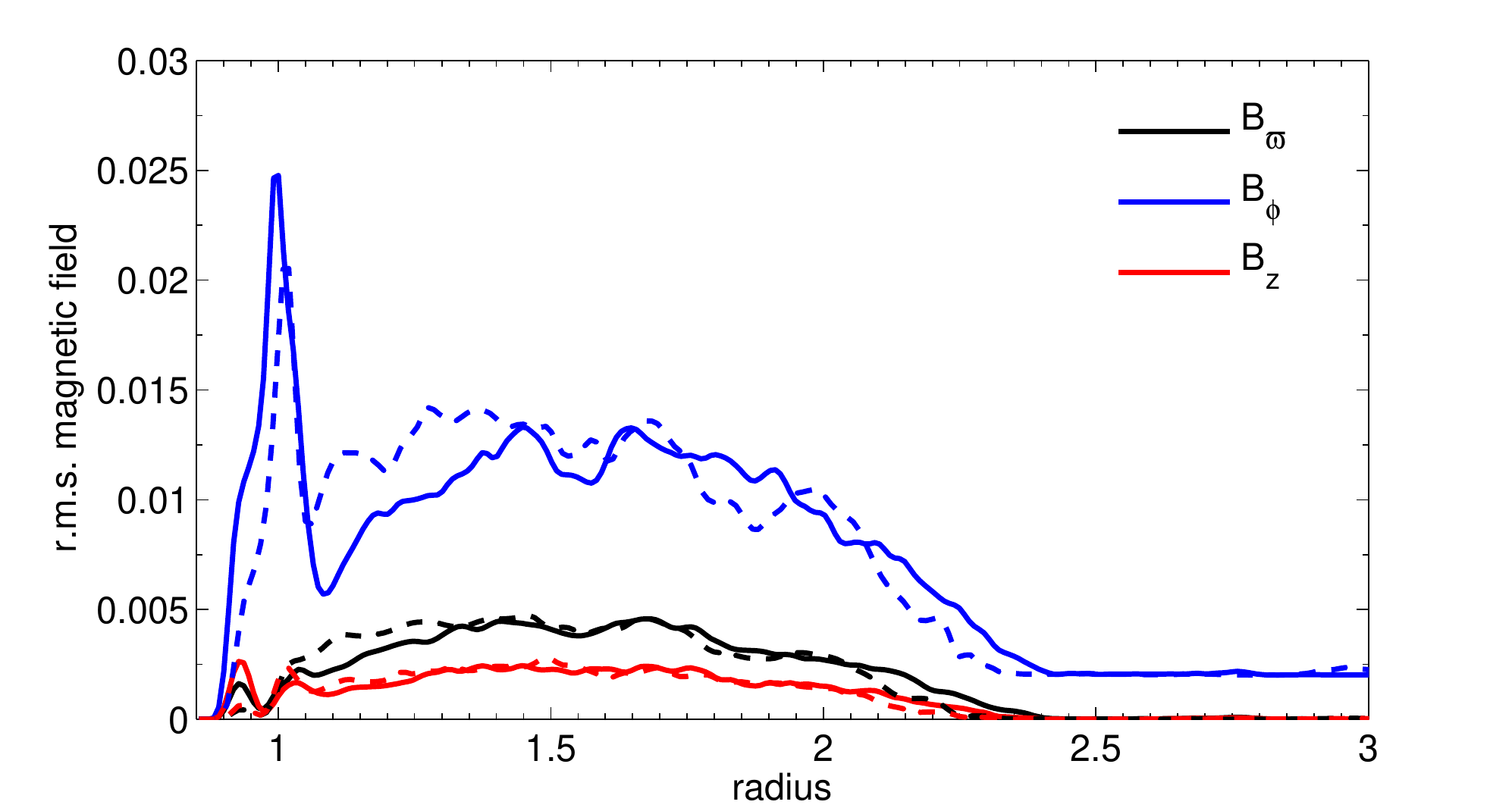}}
\subfigure[]{\includegraphics[width=0.49\textwidth]{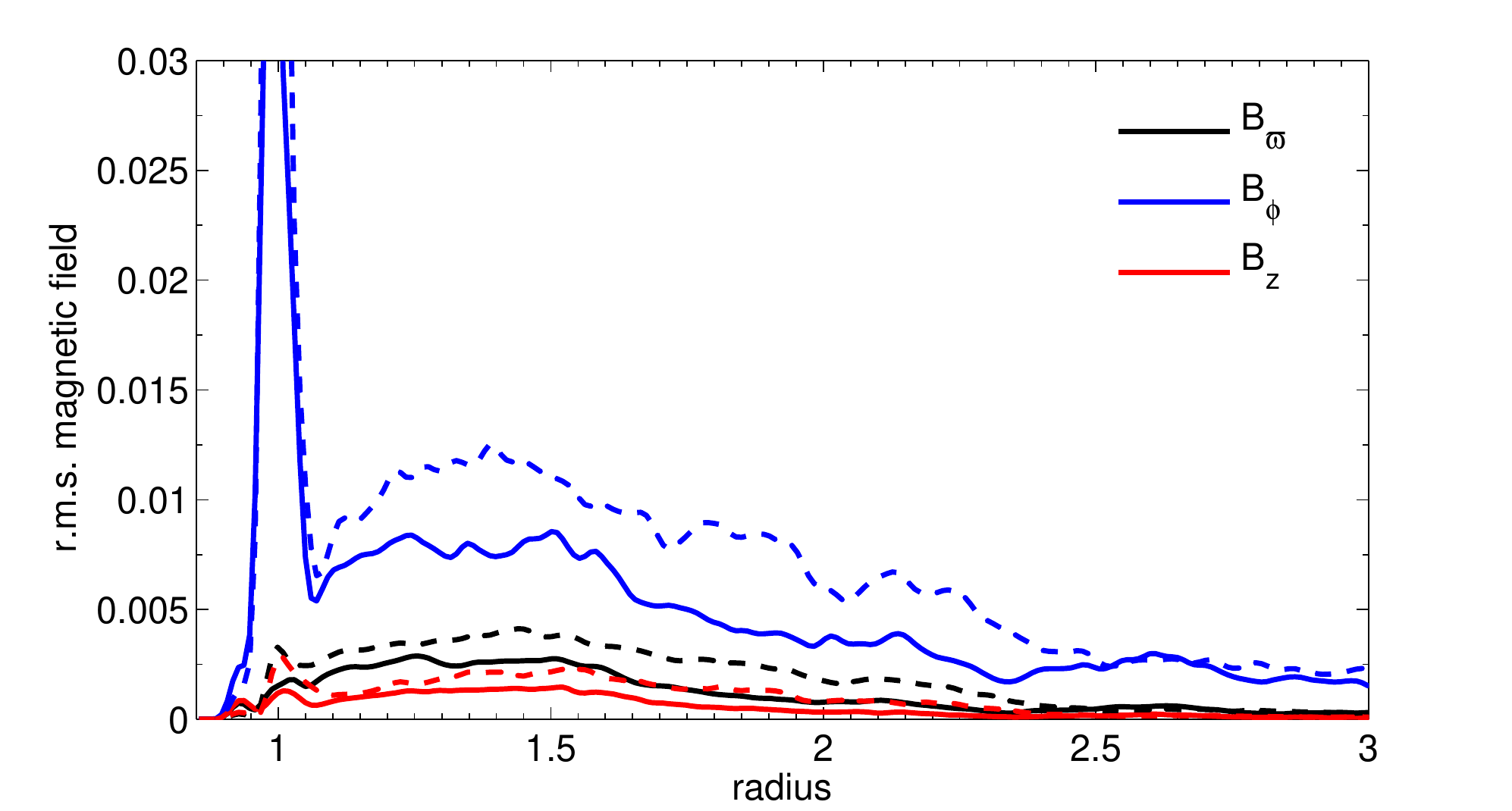}}
\caption{The volume weighted r.m.s magnetic field components at $t=400$ 
  for NVF simulations (panel
  a), NAF simulations (panel b), and ZNF simulations (panel c). In
  each case, the solid line is the simulation with the higher
  z-resolution.}
\label{Bconvplot}
\end{figure}

The second diagnostic of convergence we perform is a measurement of
the r.m.s. volume weighted average
of $B_\cp$, $B_\phi$, and $B_z$ for
simulations with the same initial magnetic field geometry, but having
a different resolution. Panels a,b,c of
Fig. \ref{Bconvplot} show r.m.s. averages of $B_\cp$, $B_\phi$, and
$B_z$ at $t=400$ for NVF simulations (M9a,d), NAF simulations (M9b,e),
and ZNF simulations (M9c,f), respectively. We see that the NVF and NAF
simulations are converged, although the magnetic field in the outer disk
of the NAF simulation is still approaching a steady state. The reason
why we did not perform the convergence study at a later time is that it
was too expensive to run the high resolution simulations beyond a time
of $t=450$.

In the ZNF case, magnetic field components for the low
resolution simulation have a higher amplitude than for the high
resolution simulation, and we have checked that
this is true at all times, not just at $t=400$. However, in (\S
\ref{alphabetasec})
we showed that MRI turbulence is resolved in ZNF simulations based on
the pitch angle indicator. 

Taken together, these two pieces of
information suggest that MRI turbulence in our ZNF simulations is
resolved, but that for a given simulation, the r.m.s field
strength in the 
saturated state of MRI turbulence is influenced by the effective numerical
viscosity and resistivity, which depend on the grid resolution
\citep{Fromang}. In order to make the results of ZNF simulations independent of
resolution, we could add explicit viscosity and resistivity \citep{Fromang2}.

However, given the exploratory nature of this paper, the exact value
of the viscous and
resistive coefficients is of secondary importance. What is more
important is that for a given value of these coefficients, MRI
turbulence is properly resolved, which is indeed the case according to
the pitch angle indicator of \S \ref{alphabetasec}.

\subsection{The Value of $\alpha$ due to MRI Turbulence in the Disk}

The most important parameter that characterizes transport by
MRI turbulence in the disk is $\alpha$. However, $\alpha$ in our
simulations is not a constant and is a function of both time and radius.
One reason for this is that the timescale for
instability is longer in the outer parts of the disk than in the inner
parts, so the inner parts reach a quasi-steady state faster. However,
even in the quasi-steady state, we find $\alpha \propto \cp
^{-3/2}$, which is not a surprising scaling, and implies a constant
value of $\alpha_\nu$ (equation
[\ref{alphaSSeq}]). The reason $\alpha$ and $\alpha_\nu$ have a
different radial dependence in simulations is that the height of
the simulation domain, $\Delta z$ is fixed, whereas the scale height
in a Keplerian disk goes as $H  = s/\Omega \propto \cp^{3/2}$ for
constant $s$. Thus, we define
\ba
\alpha_\text{eff} \equiv \alpha \cp^{3/2}.
\ea
Since, $s/\Omega = \Delta z$
at $\cp = \cp_\star = 1$, $\alpha_\text{eff}$ can be thought of as the
``effective'' value of $\alpha$ for a
Keplerian disk, based on our simulation results. The assumption here
is that the angular momentum
transport rate is proportional to the vertical scale of the problem,
which is fixed for our simulations but goes as $H \propto \cp^{3/2}$
for a Keplerian disk with a constant value of $s$.

Another complication to measuring the $\alpha$ due to MRI turbulence is the
presence of waves
emitted from the BL, which contribute to the value of $\alpha$ in the
disk. Therefore, rather than showing $\alpha_\text{eff}$ directly,
we show $\alpha_\text{B,eff}$, which is simply the component of
$\alpha_\text{eff}$ that is due to magnetic stresses
(i.e. $\alpha_\text{B,eff} =
\alpha_B \cp^{3/2}$). As mentioned in \S \ref{alphabetasec},
$\alpha_\text{B}$ is insensitive to waves, which contribute
primarily to the hydrodynamical stress. Studies of the MRI
typically find that $\alpha \approx 5/4\alpha_B$ \citep{Sorathia},
which also roughly holds for our simulations.

\begin{figure}[!h]
\centering
\subfigure{\includegraphics[width=0.8\textwidth]{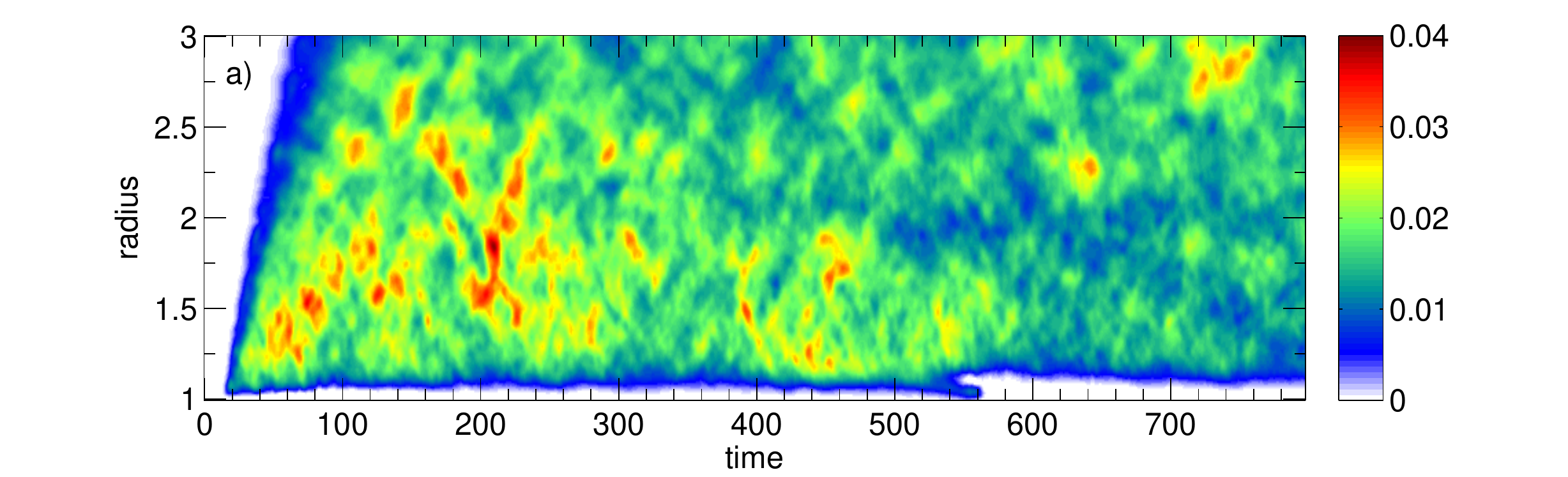}}
\subfigure{\includegraphics[width=0.8\textwidth]{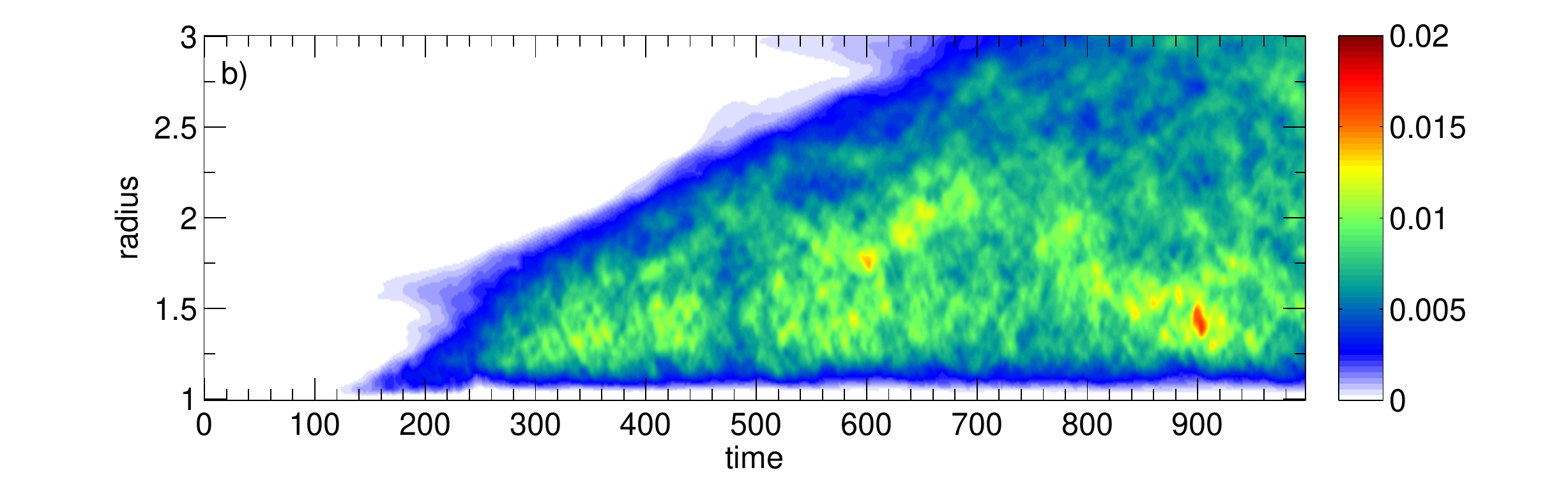}}
\subfigure{\includegraphics[width=0.8\textwidth]{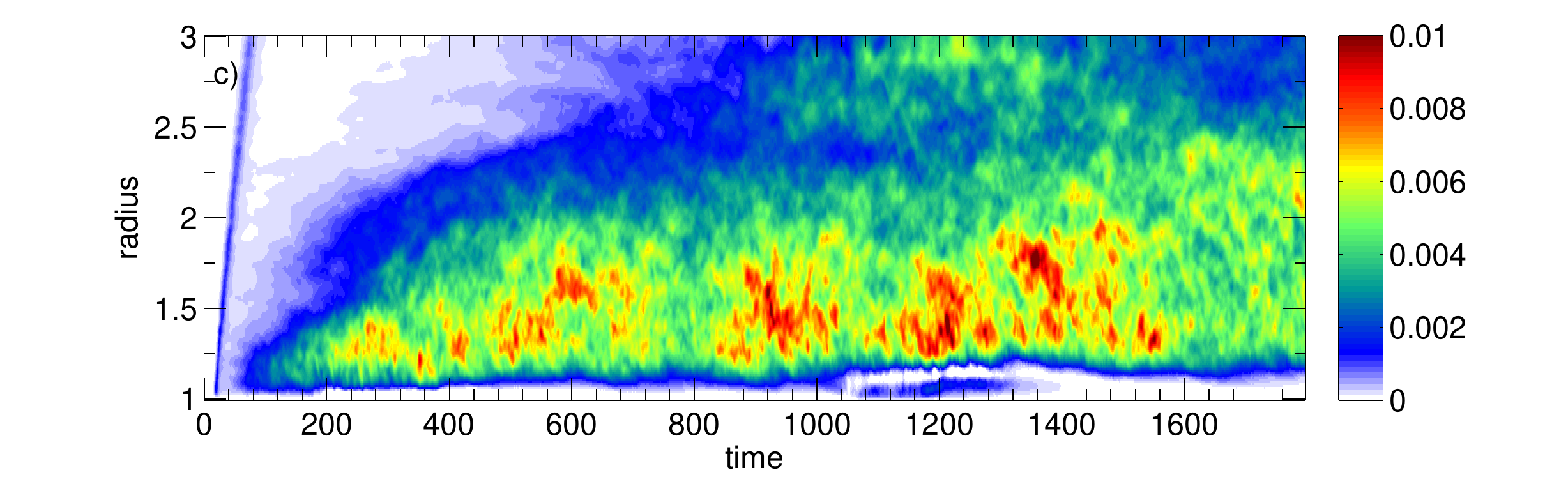}}
\caption{Panels a,b,c show $\alpha_\text{B,eff}$ for simulations
  M9d,e,f respectively. Note the different scale on the time axis in
  the figures.}
\label{alphaBspacetime}
\end{figure}

Fig. \ref{alphaBspacetime} shows radius time plots of $\alpha_\text{B,eff}$ for
simulations M9d,e,f. Once a quasi-steady state has been
reached, the value of $\alpha_\text{B,eff}$ for the
NVF simulation (M9d) is $\alpha_\text{B,eff} \approx .02-.03$, for the
NAF simulation (M9e) is 
$\alpha_\text{B,eff} \approx .0075-.0125$, and for the ZNF simulation (M9f) is
$\alpha_\text{B,eff} \approx .004 - .01$. These values of $\alpha$ are
consistent with those obtained by \citet{SteinackerPapaloizou} in
their BL simulations. However, the value of $\alpha$ in ZNF
simulations does depend on the resolution, for the same reasons as the
r.m.s. magnetic field value depends on the resolution (\S \ref{convergesec}).

\section{Magnetic Field Amplification in the BL}
\label{sect:Bampl}

It has been speculated that the
strong shear present in the BL can amplify the magnetic field
to the point that $\beta < 1$ \citep{Pringle1989}. \citet{Armitage}
observed amplification of the field in the BL in MHD
simulations with an isothermal equation of state, but found
$\beta^{-1} \lesssim .1-.2$, everywhere throughout the simulation domain,
including the BL. His setup assumed initial net vertical field 
and is most similar to our NVF model.

We also observe amplification of the $B_\phi$ field component in the BL 
in our simulations, although the amplification is highly
time-variable. Referring back to Fig. \ref{Bconvplot}, we see a bump
in $B_\phi$ within the BL for the ZNF case (panel c), which clearly
indicates shear amplification of the $B_\phi$ component. There are
also bumps in $B_z$ and $B_\phi$ in panels a and b of
Fig. \ref{Bconvplot}, respectively. However, since panels a and b
correspond to simulations
having net vertical and azimuthal flux, respectively, the bump in
$B_z$ and $B_\phi$ within the BL is due, at least in
part, to a combination of flux freezing and compression of accreted material
from the disk on the surface of the star (\S \ref{gapsec}). 

\begin{figure}[!h]
\centering
\includegraphics[width=0.7\textwidth]{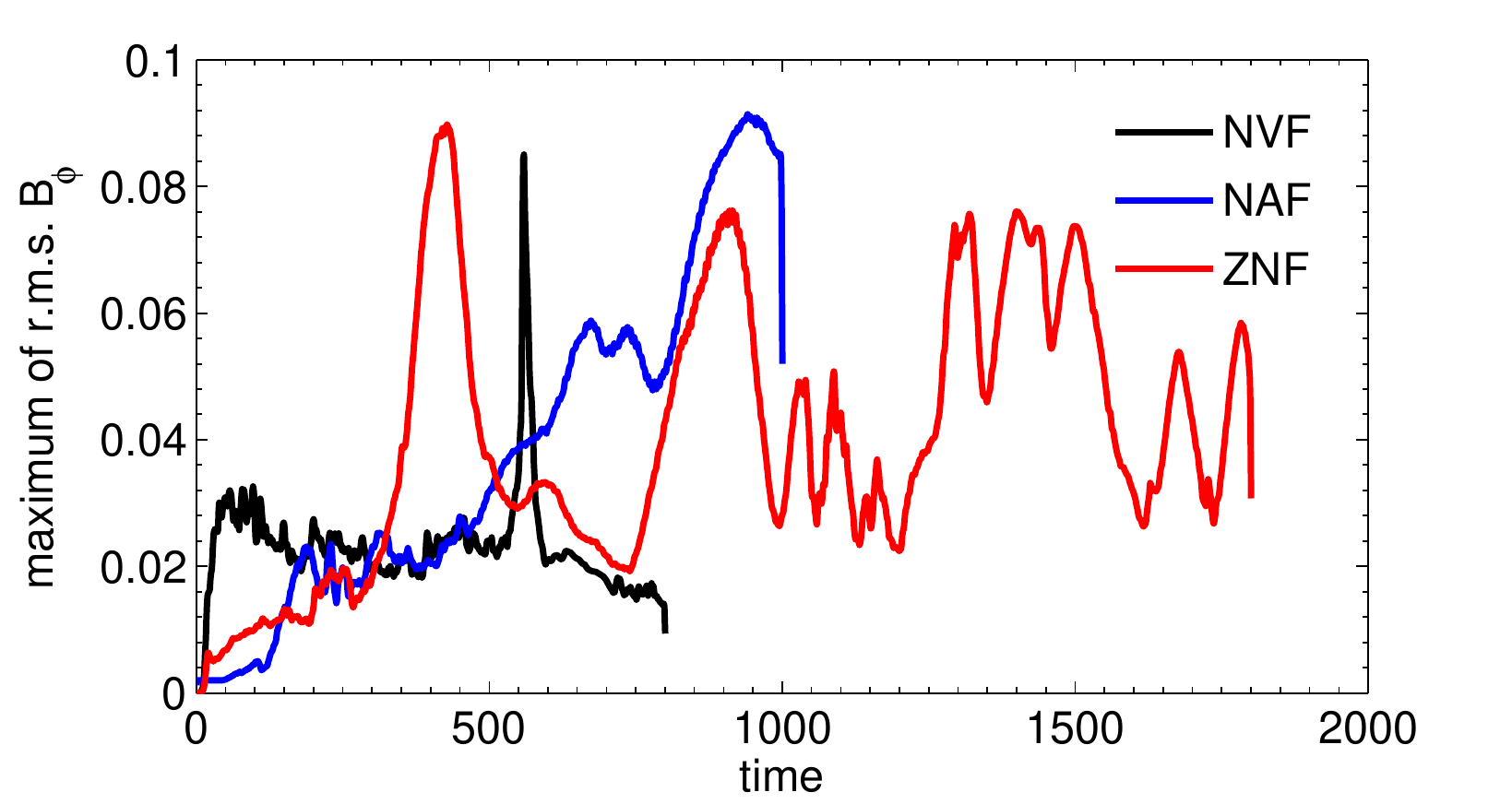}
\caption{Plot of the maximum (in $\cp$) of the r.m.s average (in $z$
  and $\phi$) of $B_\phi$ as a function of time for simulations
  M9d,e,f (NVF, black curve; NAF, blue curve; ZNF, red curve). The
  spikes correspond to transient shear amplification of the $B_\phi$
  component of the field within the BL.}
\label{maxrmsfig}
\end{figure}

To disentangle amplification of magnetic field due to
flux freezing and compression from shear amplification, we
plot the maximum (in $\cp$) of the r.m.s averaged field (in $z$ and $\phi$) as
a function of time in Fig. \ref{maxrmsfig} for simulations M9d,e,f. We
see that the NVF and
ZNF simulations (black and red curves) exhibit peaks, and the
NVF simulation (black curve), in particular, has one large peak at $t
\approx 560$. The NAF simulation (blue curve), on the other hand,
exhibits a steady increase and contains small local peaks. The steady
increase can be
understood as compression of advected azimuthal field within the BL. 

\begin{figure}[!h]
\centering
\includegraphics[width=0.7\textwidth]{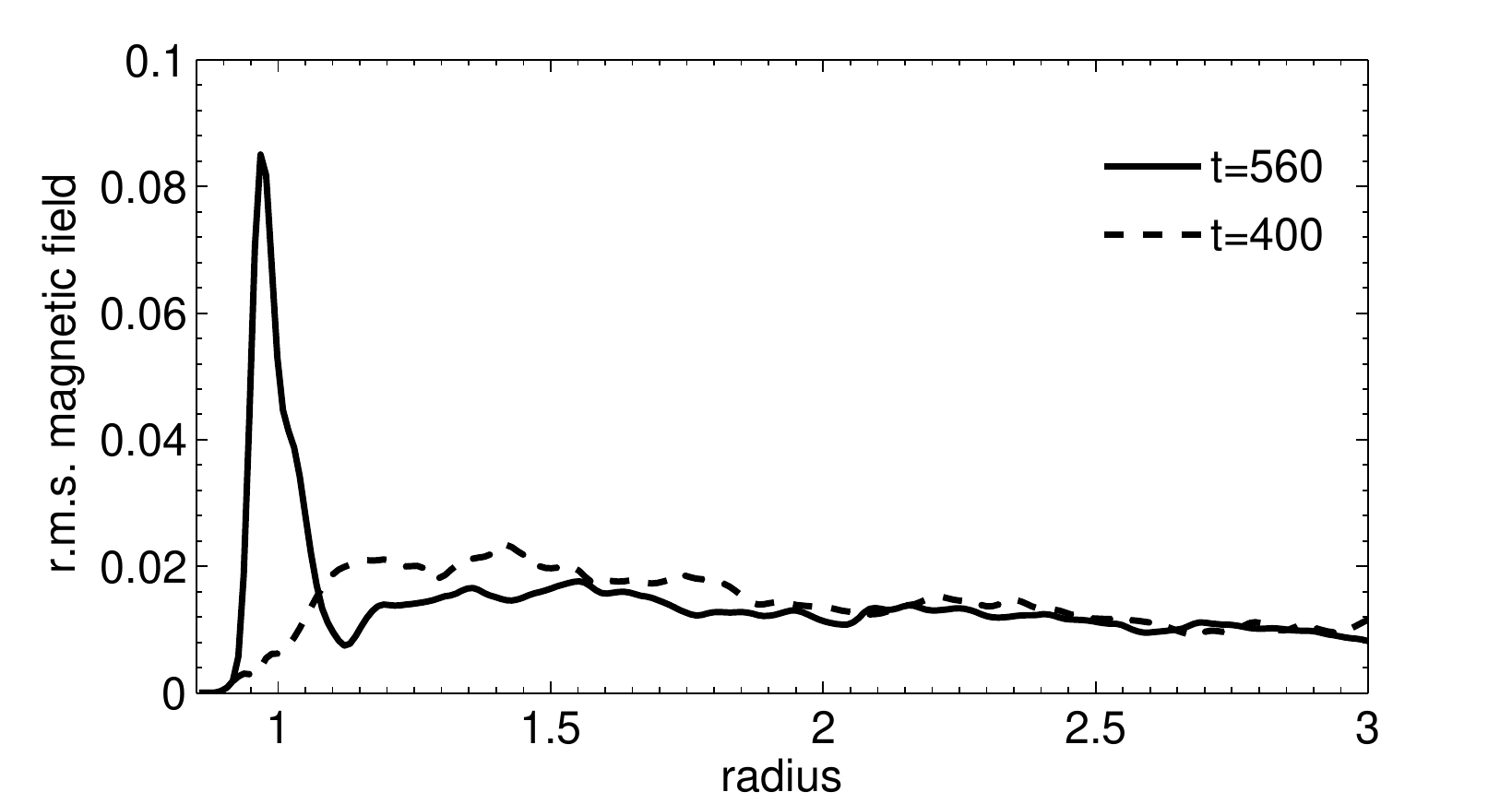}
\caption{Plot of the r.m.s average (in $z$ and $\phi$) of $B_\phi$ for
simulation M9d (NVF) at two different points in time. The dashed line
corresponds to $t=400$ when there is little shear amplification of the
magnetic field in the BL and the solid line to $t=560$ when shear
amplification reaches a maximum in time.}
\label{Bphihilo}
\end{figure}

The peaks in Fig. \ref{maxrmsfig} are due to transient shear amplification of
the field in the BL, and the baseline level in the NVF
simulations (black curve) is simply set by the value of $B_\phi$ in
the saturated state of MRI turbulence in the disk. Fig. \ref{Bphihilo}
shows the r.m.s value of $B_\phi$ as a function of radius from
simulation M9d at times $t=400$ when there is little or no
amplification and $t=560$, which corresponds to the peak in the black
curve in Fig. \ref{maxrmsfig}. From Fig. \ref{Bphihilo} it is clear
that amplification of the field does indeed take place in the BL and
that it is time-dependent. The cause for the time-dependency has
not been explored, but it is possible that it is related to the
re-excitation of acoustic modes within the BL (\S \ref{angmomwave}).

\begin{figure}[!h]
\centering
\subfigure{\includegraphics[width=0.8\textwidth]{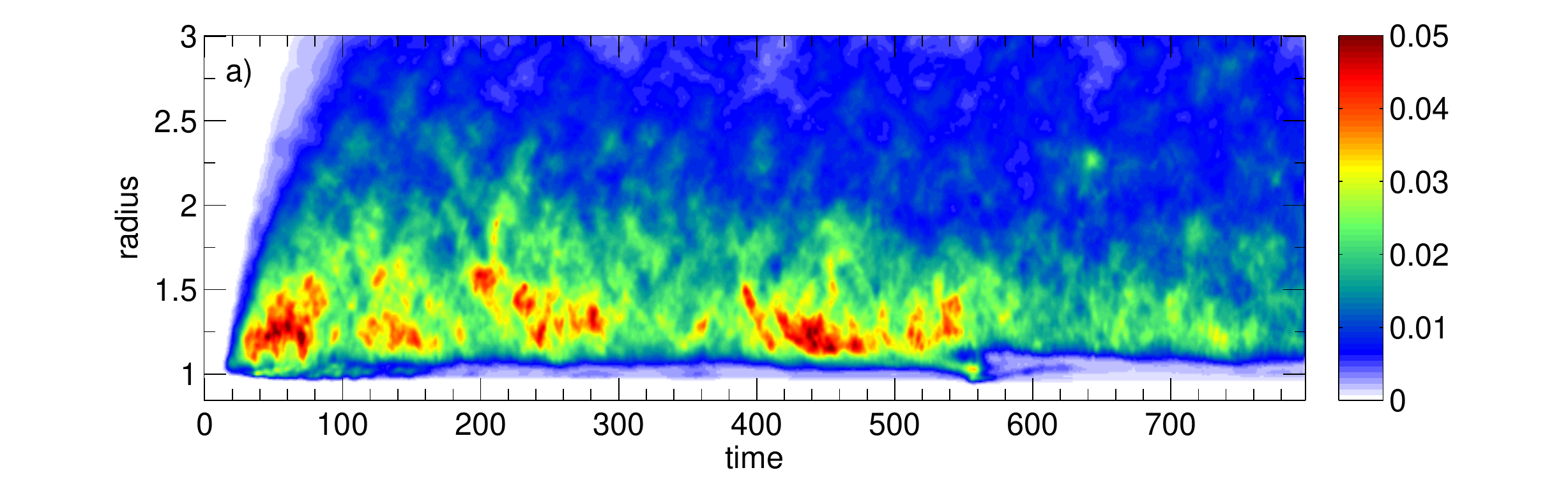}}
\subfigure{\includegraphics[width=0.8\textwidth]{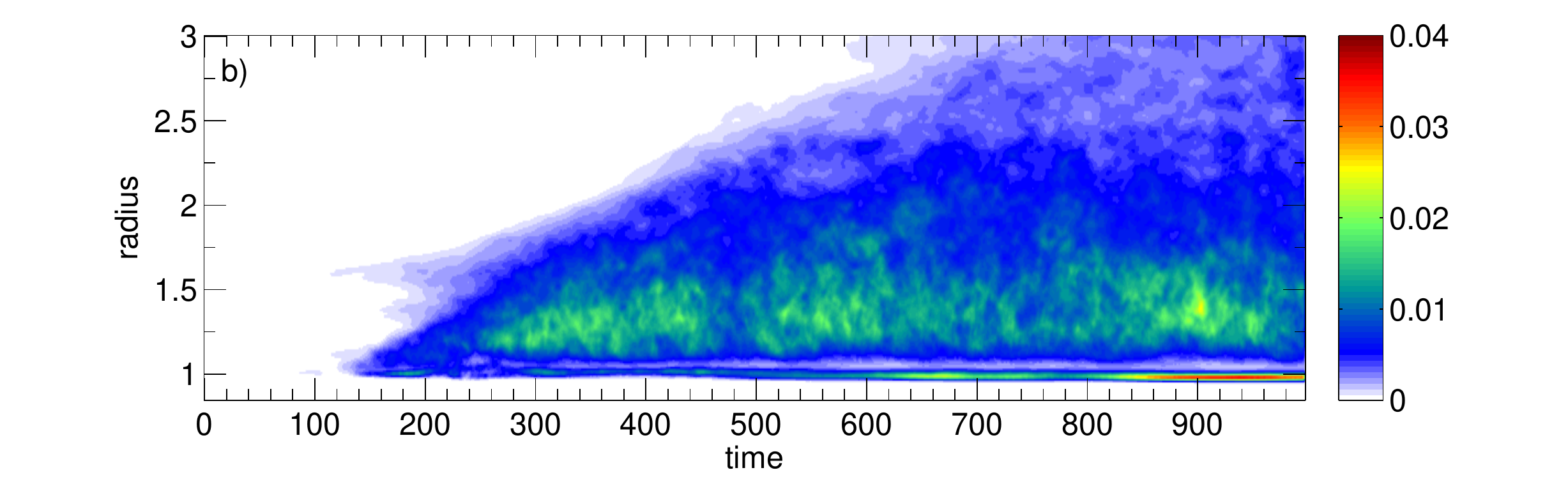}}
\subfigure{\includegraphics[width=0.8\textwidth]{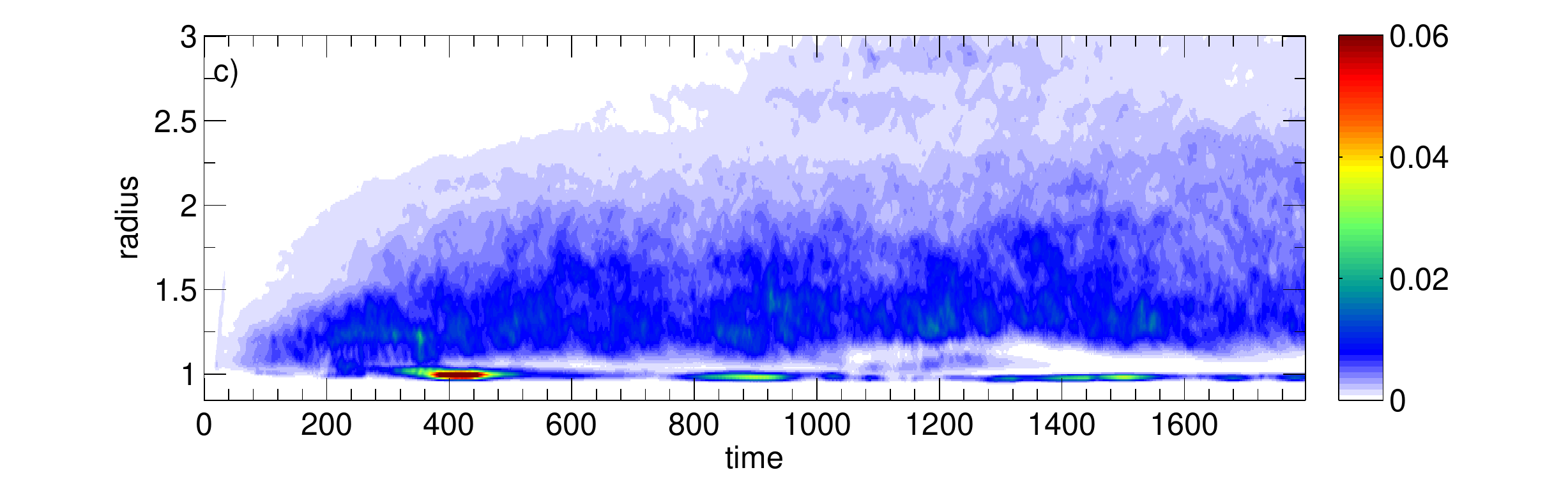}}
\caption{Panels a,b,c show radius time plots of $\beta^{-1}$
  from simulations M9d,e,f respectively. There is evidence for
  amplification of the magnetic field in the BL, but the field remains
  strongly subthermal.}
\label{betaspacetime}
\end{figure}

Although there is field amplification in the BL in our simulations,
the field always remains strongly subthermal, consistent with the
findings of \citet{Armitage}. Fig. \ref{betaspacetime}
shows radius-time plots of $\beta^{-1}$ (equation [\ref{betadef}]) for
simulations M9d,e,f, respectively. If magnetic pressure and thermal
pressure are comparable, then $\beta \sim 1$ and if thermal pressure
dominates magnetic pressure, $\beta \gg 1$. In the figures,
$\beta^{-1} \lesssim
.06$ for all time and throughout the entire simulation domain, which implies a
significantly subthermal field.

\section{Acoustic Modes in the MHD context}
\label{acoussec}

Before beginning a discussion of angular momentum transport by waves
in the BL and the star, we first show that waves are indeed present in
our simulations. The waves being referred to are the MHD analogs of
the acoustic
modes discussed by \citet{BRS,BRS1} in the hydrodynamical case. We
start by discussing how the hydrodynamical dispersion relation is
modified in the presence of a magnetic field and then discuss
simulation results.

\subsection{Modification to the Hydrodynamical Dispersion Relation}
\label{disrelsec}

\citet{BRS1} found that for the isothermal BL, there
were three branches of acoustic modes, and they found dispersion relations that
matched simulation results for each of these three branches. However,
their results were for the pure hydro case, and we comment on how they
should be modified to account for the presence of a magnetic field in
the disk. 

It is natural to expect that the
major modification to the dispersion relation of the acoustic modes in
the limit of $\beta \gg 1$ is the replacement of the sound speed with
the magnetosonic speed.
\ba
v_\text{ms} &=& \sqrt{s^2 + v_A^2} \\
\label{vms2}
&=& s \sqrt{1 + \frac{2}{\gamma} \beta^{-1}}
\ea
where $v_A$ is the Alfv\'en speed, and $\gamma = 1$ for an isothermal
equation of state.

In reality, the situation is slightly more complicated,
and the dispersion relation depends on the jump in magnetic field
across the BL. We show this explicitly in Appendix \ref{disapp}, where
we derive an analytic dispersion
relation for the 2D modes of a compressible vortex sheet
\citep{MilesKH,Gerwin,BR} in the presence of a magnetic
field that is perpendicular to the plane of the modes. This is not a
realistic setup, but it highlights how the hydrodynamical
dispersion is modified when a magnetic field is introduced. 

The main
result of the analysis, in the context of astrophysical BLs, is that
in the limit of $\beta \gg 1$, the dispersion relation for the model
system is only modified by terms of $\mathcal{O} \left( \beta^{-1}
\right)$. This result, which may be expected to hold for more general
field configurations, shows that when gas pressure dominates magnetic pressure,
acoustic modes should still be present, and their properties should
only be slightly modified as compared to the hydro case.
In a typical astrophysical context, such as in MRI turbulence, the
magnetic field is subthermal, meaning $\beta > 1$. As we have shown 
in \S \ref{sect:Bampl}, the same is also true inside the BL, meaning
that the modes present in the layer and its vicinity are essentially
acoustic modes, weakly modified by a magnetic field.

In addition to magnetosonic waves, an ideal fluid in the presence
of a magnetic field also supports Alfv\'en and slow waves. One may
wonder whether there are also modes of the system that correspond to
a coupling of these waves in the disk across the BL to either
gravity or sound waves in the star. This subject is
beyond the scope of the present work. However, we mention that in our
simulations, the modes that show up most clearly are the analogs of
the acoustic modes discussed by \citet{BRS,BRS1}. Thus, it is natural
to expect that these magnetosonic modes are the most
important ones for transporting angular momentum in the presence of a
magnetic field, when $\beta \gg 1$. 

\subsection{Simulation Results} 

\begin{figure}[!h]
\centering
\subfigure[]{\includegraphics[width=0.55\textwidth]{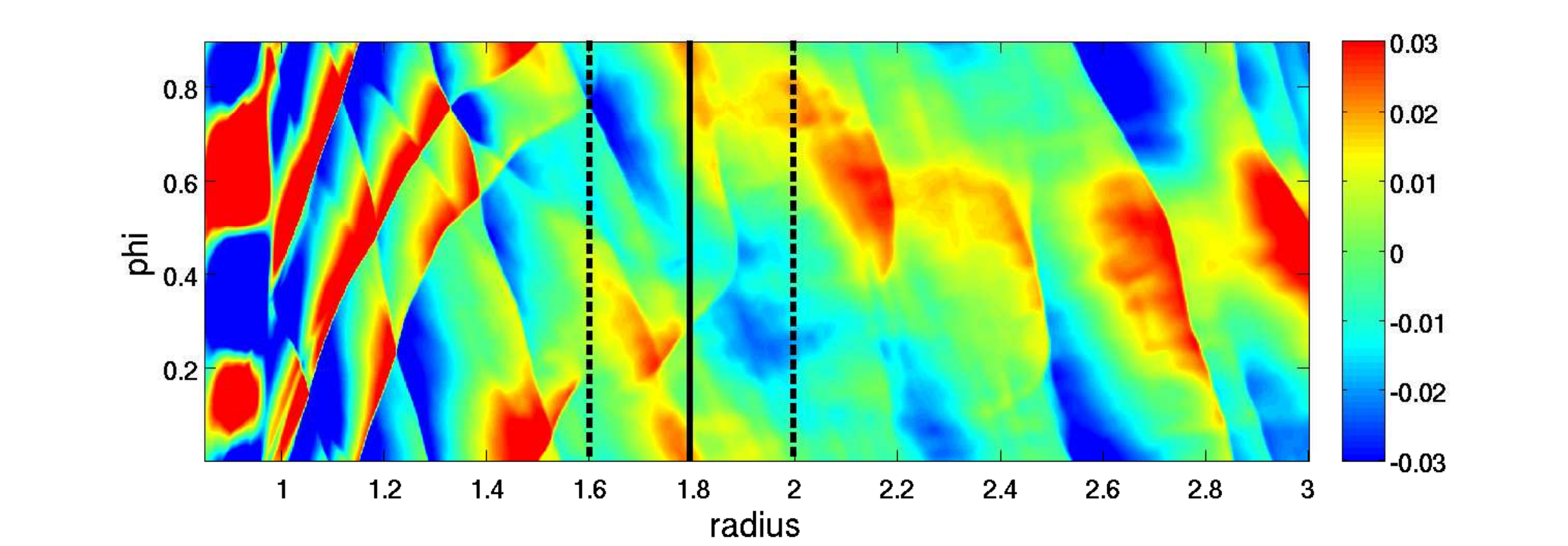}}
\subfigure[]{\includegraphics[width=0.55\textwidth]{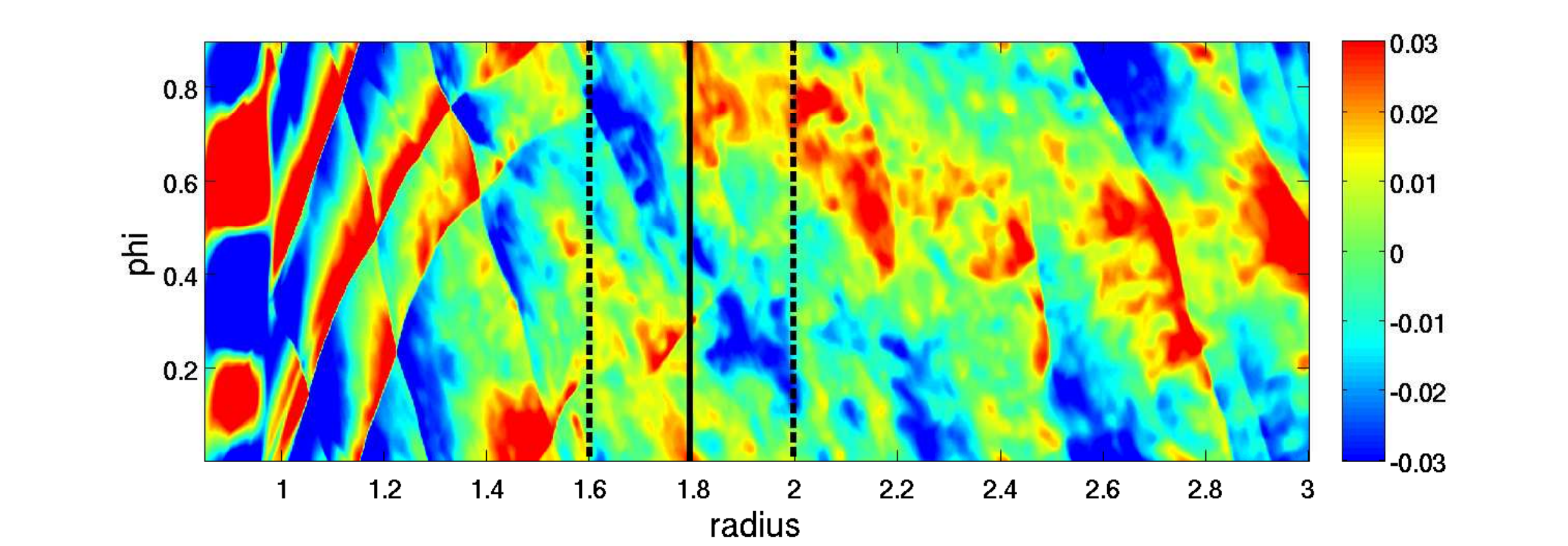}}
\subfigure[]{\includegraphics[width=0.55\textwidth]{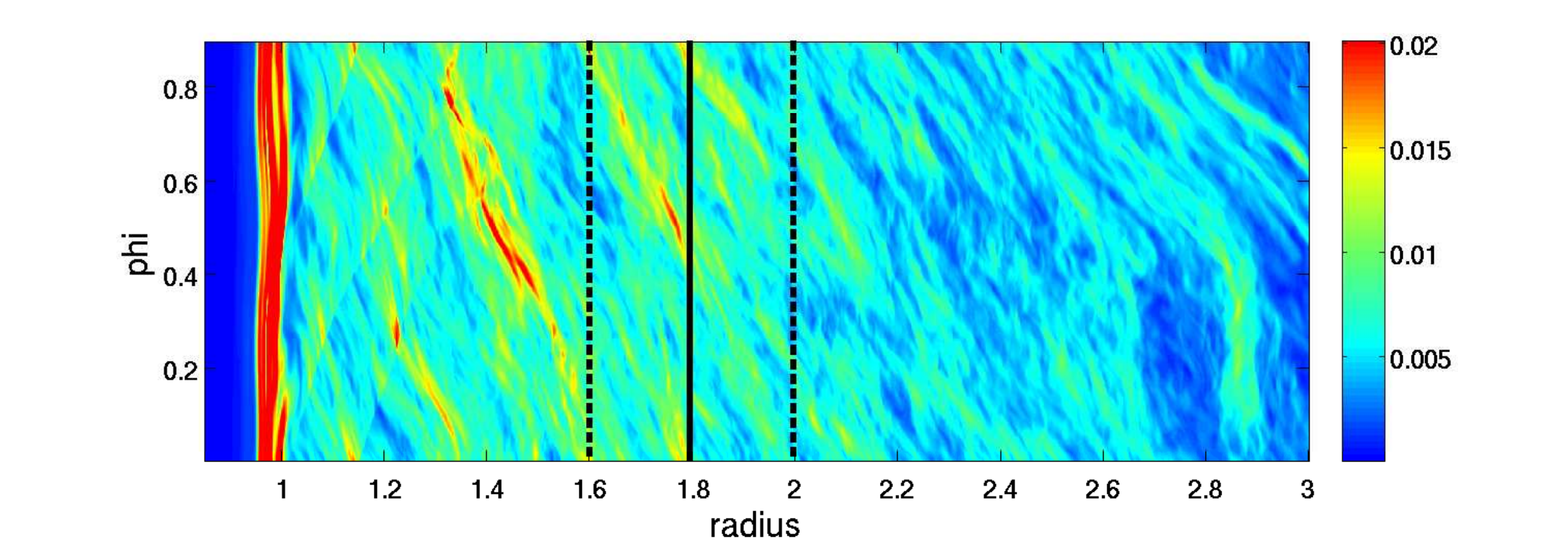}}
\subfigure[]{\includegraphics[width=0.55\textwidth]{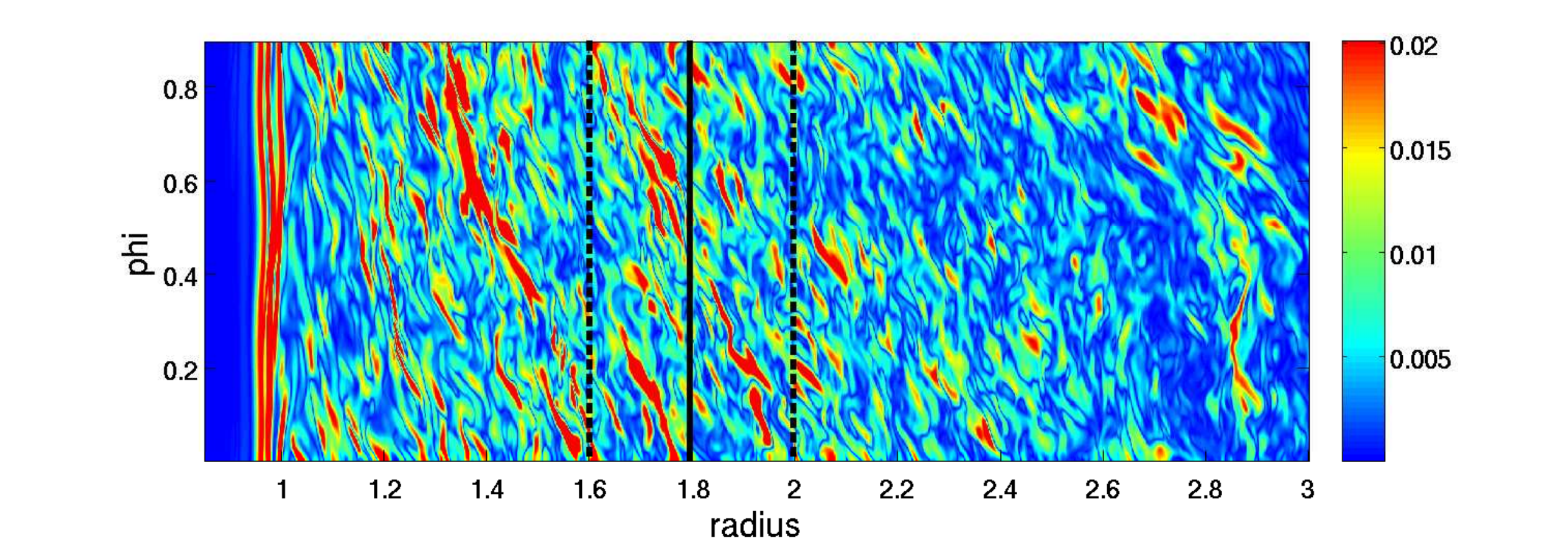}}
\caption{Snapshots at $t=1000$ from simulation M9f. Panel a) shows
  the simple vertical average of $\cp \sqrt{\rho} v_\cp $, and b) shows
  a slice through the $z = 0$ plane of $\cp \sqrt{\rho} v_\cp $. Panel c)
  shows the simple vertical average of the magnetic field magnitude
  and d) shows a slice through the $z=0$ plane of the magnetic
  field magnitude. The dashed lines mark the edges of the evanescent
  region for the $m=14$ lower branch mode, and the solid line marks
  the location of the corotation radius in the disk for that mode.}
\label{znf_snapshot}
\end{figure}

We observe excitation of magnetosonic modes in our simulations, which
persist, typically, for
the duration of the simulation. Fig. \ref{znf_snapshot} shows both 
the simple vertical average and the cross-section through the $z=0$ plane of
both $\cp \sqrt{\rho} v_\cp$ and the magnetic field magnitude for
simulation M9f at time $t=1000$. The dominant mode in
Fig. \ref{znf_snapshot} is the $m=14$ lower
branch mode\footnote{For a discussion of wave branches, including the lower
branch, see \citet{BRS1}.}.
It appears prominently in  $\cp \sqrt{\rho} v_\cp$ (panels a and
b), but only weakly in the magnetic field magnitude (panels c and d). The
measured pattern speed of the $m=14$ mode from simulations is 
$\Omega_P \approx .4$, which is in good agreement with the theoretically
predicted pattern speed for the $m=14$ lower branch mode, $\Omega_P =
.415$ \citep{BRS1}. 

We also point out that the mode depicted in
Fig. \ref{znf_snapshot} is effectively two-dimensional, since there is
not much difference between the vertically averaged image of $\cp
\sqrt{\rho} v_\cp$ (panel a) 
and the cross-sectional slice (panel b). The magnetic field, on the
other hand, exhibits much more vertical structure (panels c and d),
which is indicative of turbulence.

The lower branch has an evanescent region in the disk around the 
corotation radius and waves incident on the evanescent region are
reflected back towards
the BL. The edges of this evanescent region are located
approximately at the Lindblad radii, which are implicitly defined by
the condition
$\Omega(\cp_\text{LR}) = \Omega_P \pm \kappa(\cp_\text{LR})/m$, where
$\kappa$ is the epicyclic frequency.
The dashed lines in Fig. \ref{znf_snapshot} show the edges of
the evanescent region, as predicted by hydrodynamical theory assuming
a Keplerian rotation profile, and the solid line shows the corotation
radius of the mode in the disk. Even though we now have MRI turbulence
in the disk, shocks still reflect off inner edge of the evanescent
region. This fact, together with the
agreement between the measured pattern speed and that from 
hydrodynamical theory, means
hydrodynamical theory provides a good description of the properties of
magnetosonic modes even in the presence of MRI turbulence in the disk. As argued
earlier, this can be attributed to the fact that $\beta^{-1} \ll 1$
in the simulations, so the magnetic field does not
significantly affect the properties of acoustic modes.

Although the hydrodynamical and MHD cases exhibit many similarities,
some differences are apparent. The first difference is that MHD simulations
are typically ``noisier'' than hydro runs \citep{BRS,BRS1} with a
superposition of modes present. The second difference is that unlike 
the hydrodynamical case,  there can be significant wave action 
tunneling through the evanescent region in MHD runs.

\begin{figure}[!h]
\centering
\includegraphics[width=0.8\textwidth]{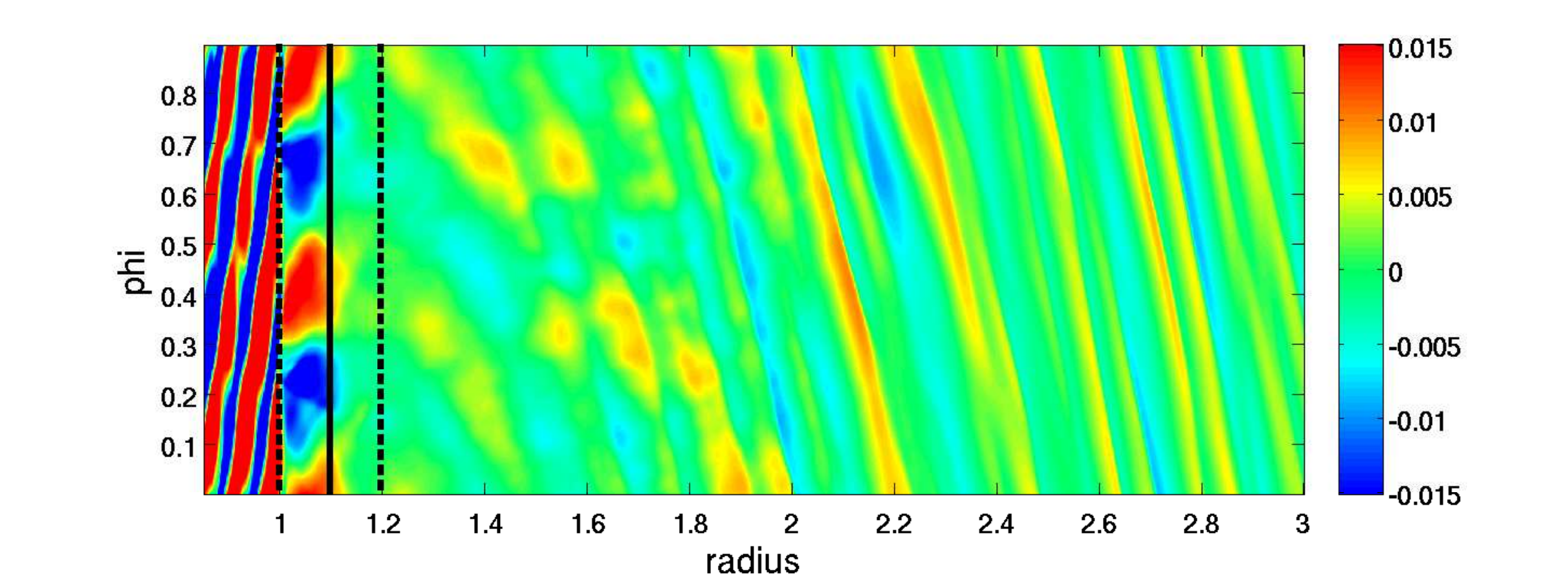}
\caption{Image of the simple vertical average of $\cp \sqrt{\rho} v_\cp$
  at $t = 40$ for simulation M9c. The $m=14$ upper branch mode is
  clearly present in
  the image. The magnetic field is not shown, since at this time, the
  MRI instability has not yet fully developed. The dashed lines mark
  the edges of the evanescent region for the $m=14$ upper mode, and the
  solid line shows the corotation radius for that mode.}
\label{upperbranchfig}
\end{figure}

The presence of a superposition of acoustic modes in the disk is
a generic feature, independent of initial magnetic field geometry. In
addition to the lower branch, which is persistent in all of our 
simulations for $t \gtrsim 60$, we also observe the upper
branch at the beginning of some simulations, but only
in transience. Fig.
\ref{upperbranchfig} shows the upper branch at $t=40$ in simulation
M9c. The mode depicted in the figure has $m=14$ and a measured pattern speed of
$\Omega_P = .84$. This agrees well with the theoretically predicted
pattern speed for this mode, which is $\Omega_P = .87$.

\section{Angular Momentum Transport by Waves in the BL, Star, and
  Inner Disk}
\label{angmomwave}

Having shown that magnetosonic modes are present in our simulations
and have similar properties to the acoustic modes discussed in
\citet{BRS1}, we now show that they are able to transport angular
momentum in the star and inner disk just as in hydrodynamical simulations. 

\subsection{Density Gap in the Inner Disk and Accretion onto the Star}
\label{gapsec}

\begin{figure}[!h]
\centering
\subfigure{\includegraphics[width=0.8\textwidth]{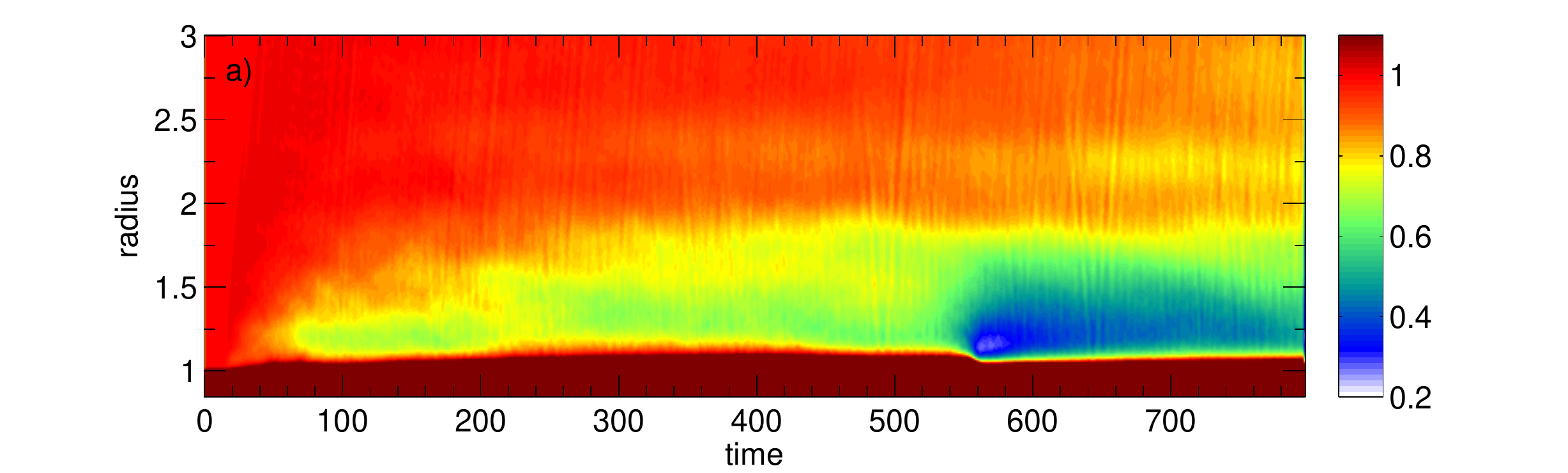}}
\subfigure{\includegraphics[width=0.8\textwidth]{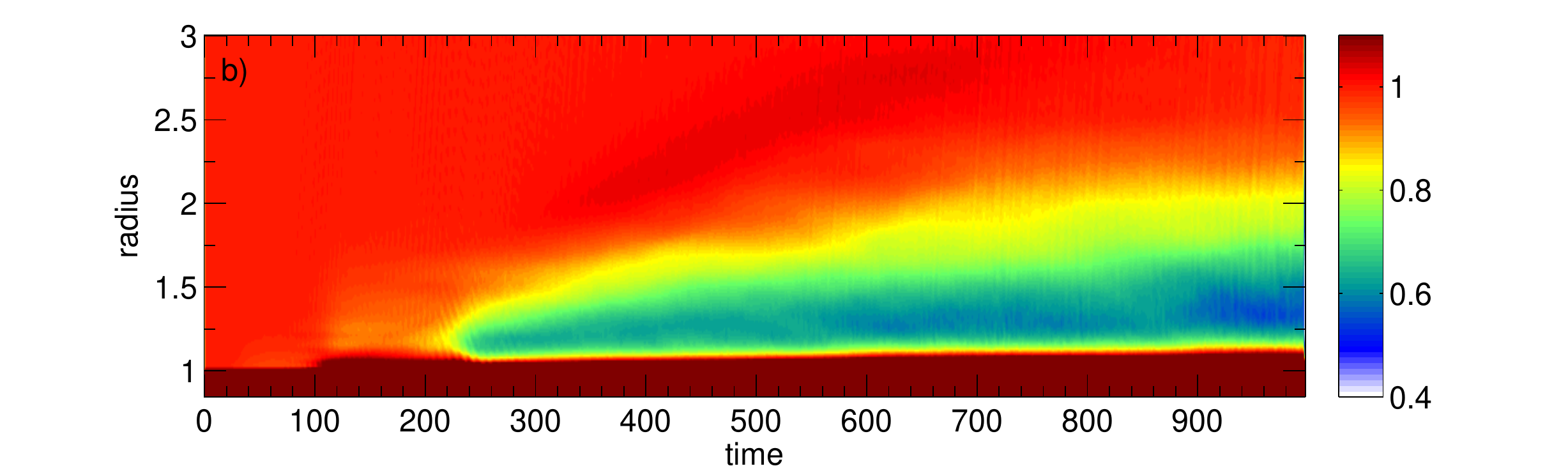}}
\subfigure{\includegraphics[width=0.8\textwidth]{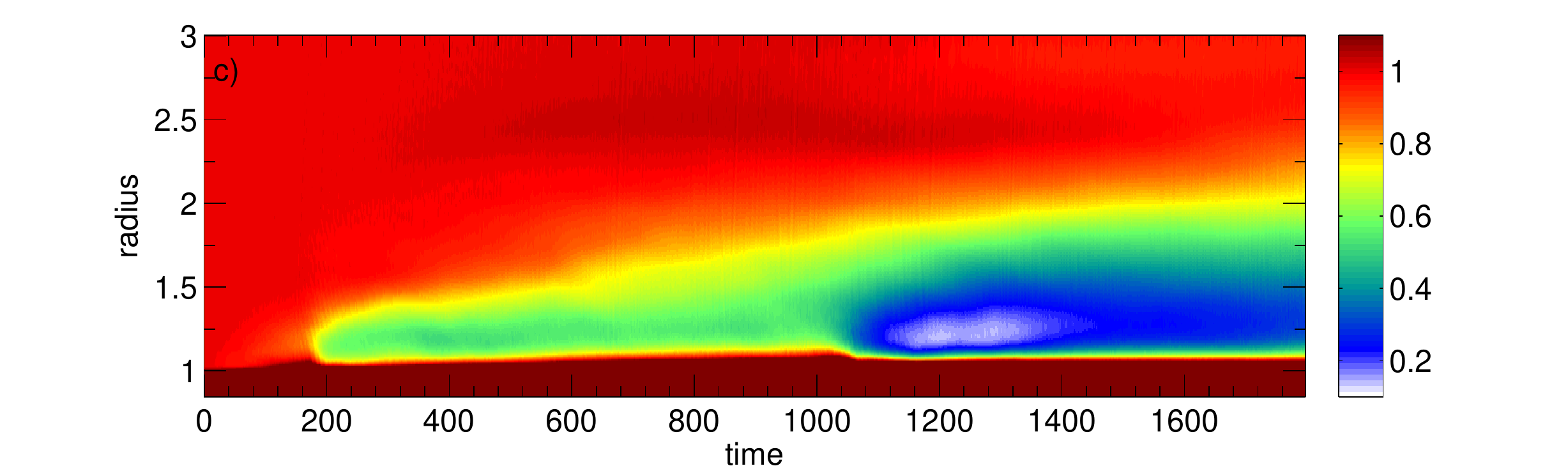}}
\caption{Panels a,b,c show radius time plots of $\Sigma_0$ from
  simulations M9d,e,f (NVF, NAF, ZNF) respectively. A density gap in
  the disk near
  $\cp = 1$ develops in all of the simulations and in panels a and c
  gap deepening is observed at a later time.}
\label{densspacetime}
\end{figure}

We begin our discussion of wave transport of angular momentum by showing
the evolution of the density profile. 
Fig. \ref{densspacetime} shows radius time plots of
$\Sigma_0(\cp,t)$ in simulations M9d,e,f. In each case, a 
gap in the density develops in the inner disk in the vicinity of the BL
($\cp \gtrsim 1$). The gap is opened up relatively rapidly, over a
time of $\lesssim 10$ orbits, and in panels a and c, gap formation is
followed at a later time by gap deepening. Concentrating on panel c, after the
initial gap
opening event at $t \approx 200$, the density profile of the gap stays
approximately constant, until $t \approx 1100$ when it undergoes
substantial gap deepening. For $t >
1100$, the gap is gradually filled in by the action of turbulent
viscosity in the disk.

\citet{BRS1} also observed gap formation in their hydrodynamical
simulations, which was driven by angular momentum transport by
acoustic modes. Consistent with their results, we find that
magnetosonic modes are responsible for opening up the gap in our MHD
simulations, and we discuss the link between acoustic modes and the
gap in the inner disk in more detail in \S \ref{angmomsec6}.
 
\begin{figure}[!h]
\centering
\subfigure[]{\includegraphics[width=0.49\textwidth]{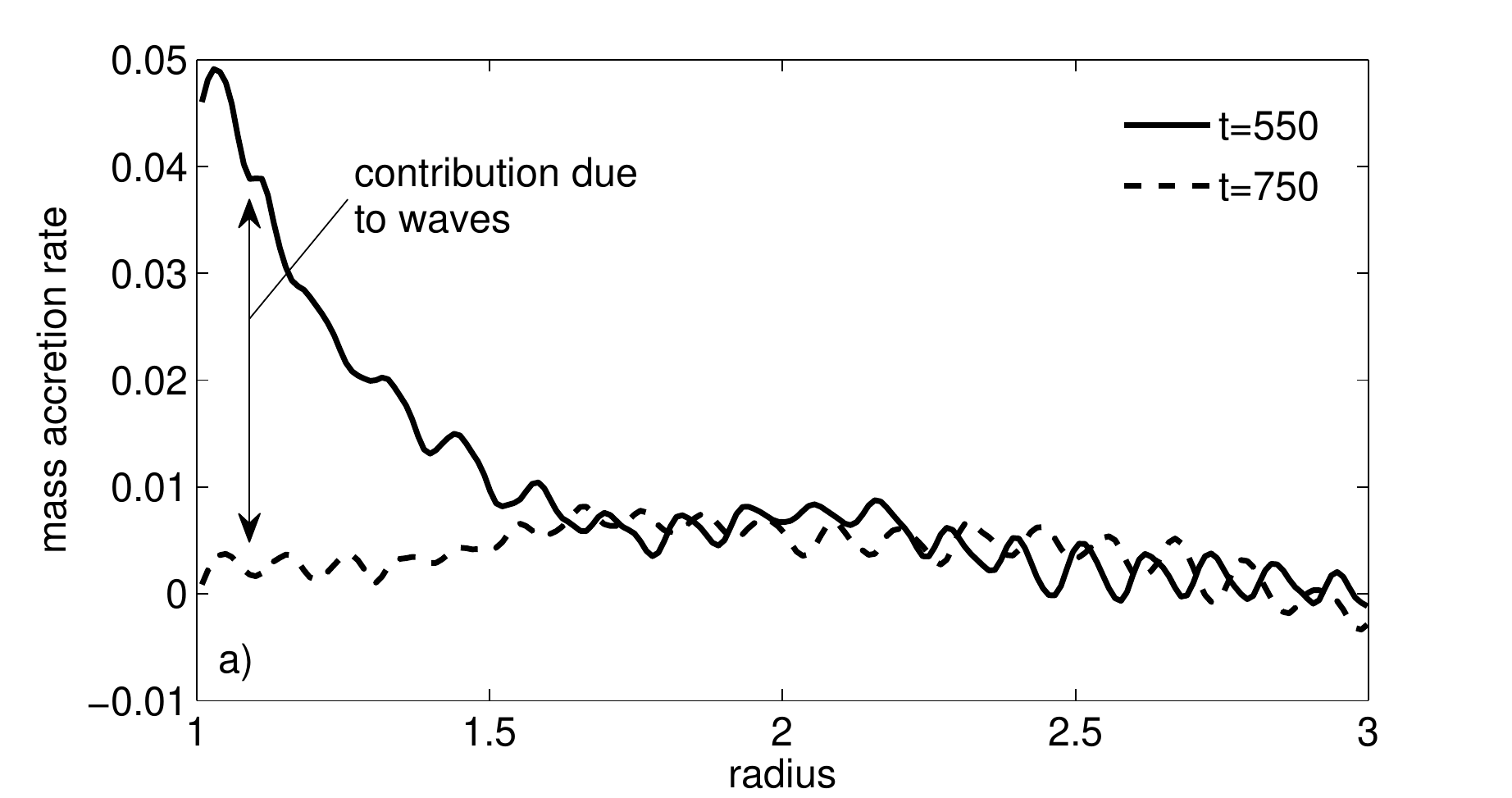}}
\subfigure[]{\includegraphics[width=0.49\textwidth]{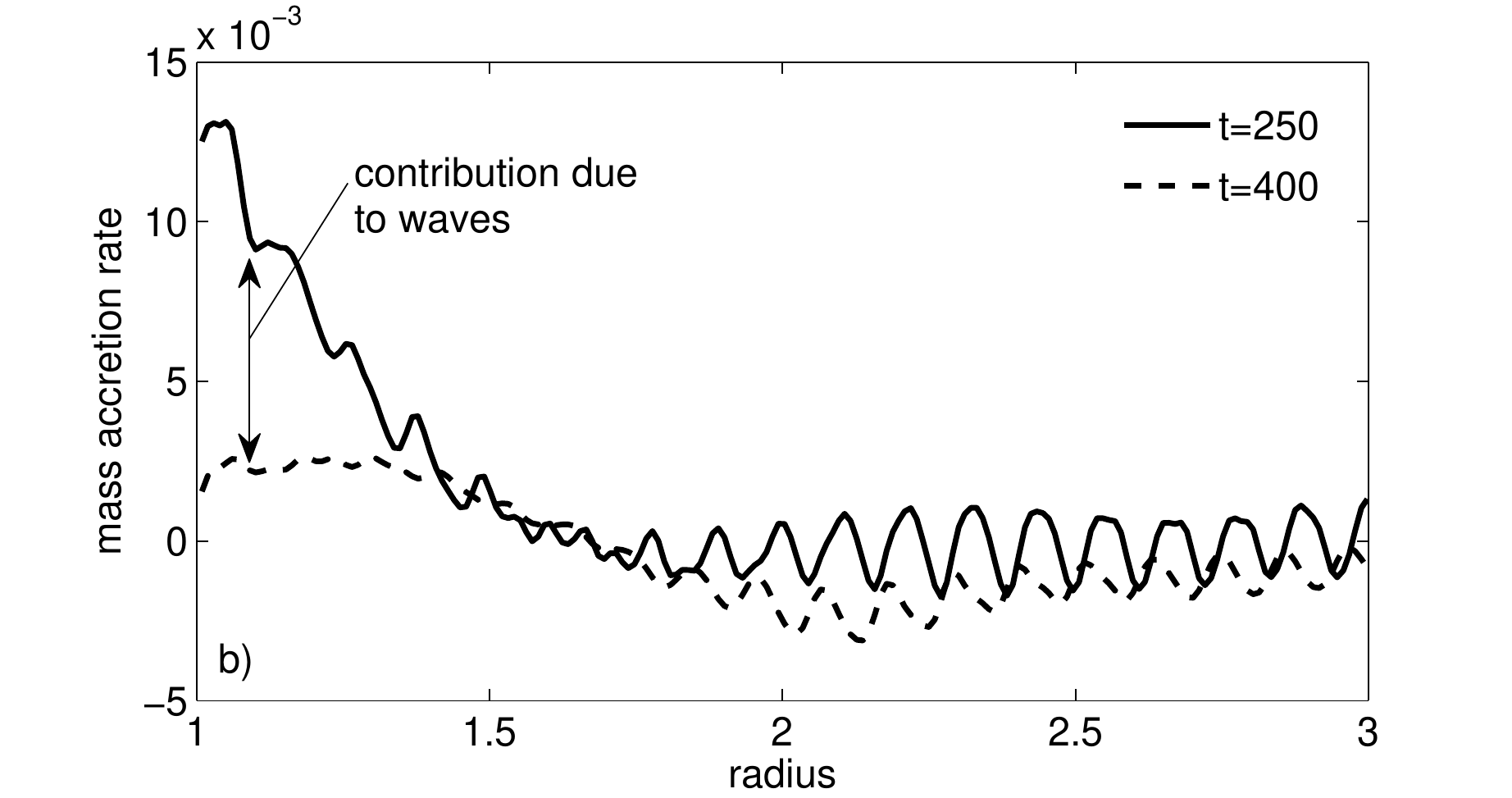}}
\subfigure[]{\includegraphics[width=0.49\textwidth]{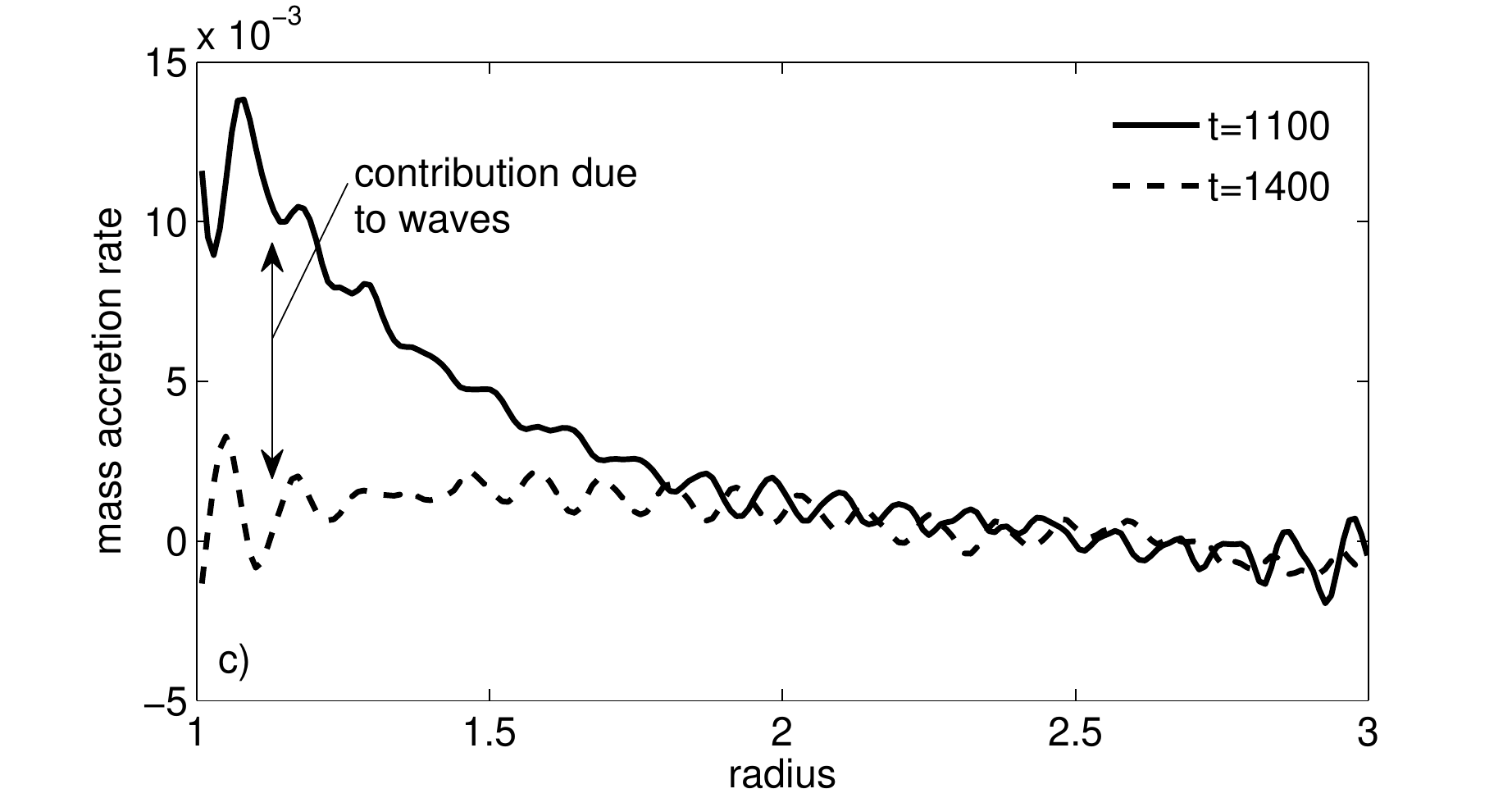}}
\subfigure[]{\includegraphics[width=0.49\textwidth]{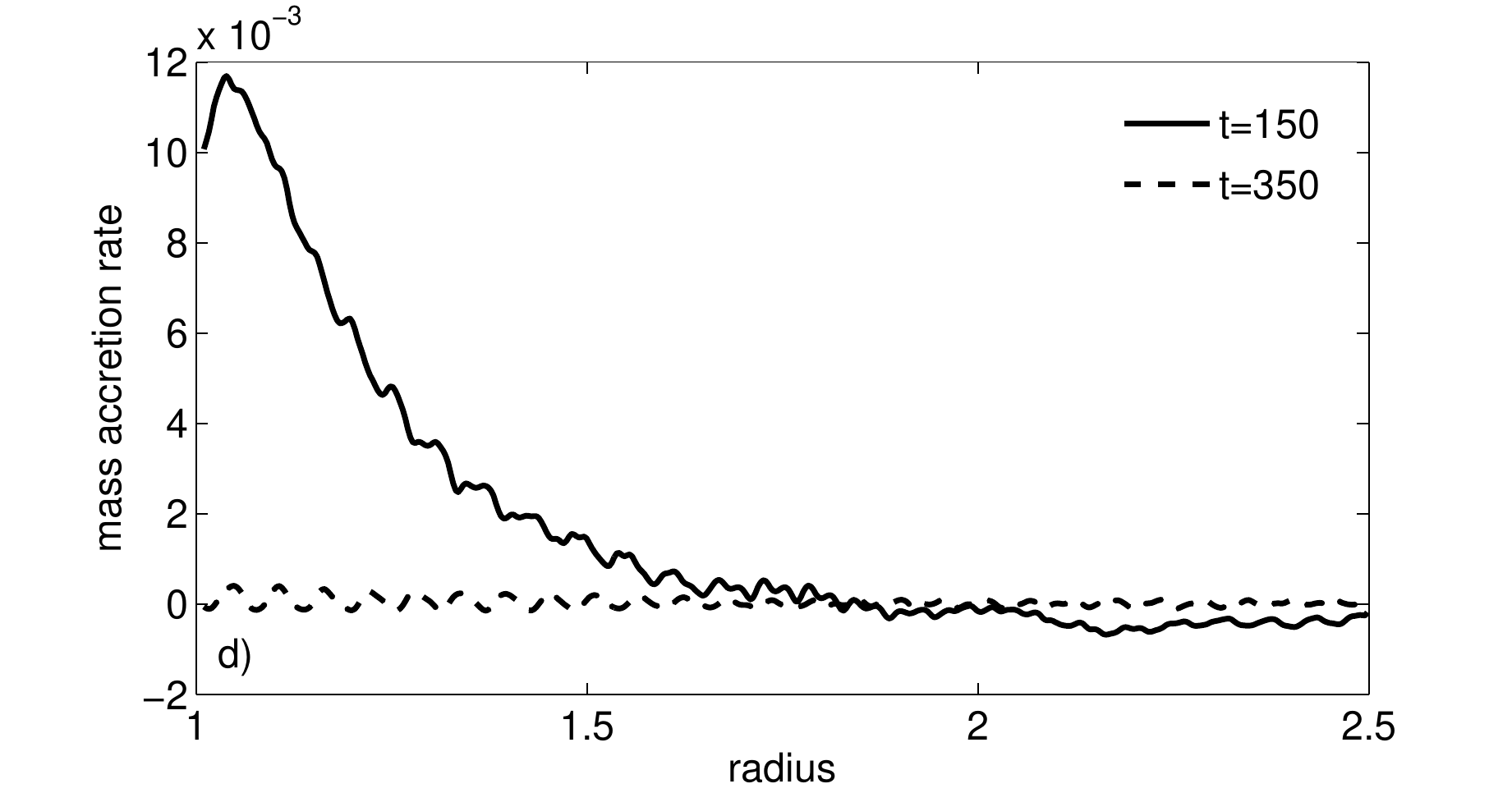}}
\caption{Panels a, b, c show plots of the instantaneous 
mass accretion rate, $\dot{M}$, from simulations
M9d,e,f (NVF, NAF, ZNF) at the times indicated in the legends.
Panel d shows the mass accretion rate
  for an $M=9$ hydro simulation considered in \citet{BRS1} (their 3D9a
  simulation). The solid and dashed lines show $\dot{M}$
during periods when acoustic modes have relatively high and low
amplitudes, respectively. The large spike in accretion rate in the
inner disk for the solid curves as compared to the dashed ones is
  caused by shock accretion in the inner disk due excitation of modes
  to a high amplitude.}
\label{mdotplots}
\end{figure}

Material gets removed from the inner part of the disk between the BL and 
the wave reflection point at the inner Lindblad resonance by the
action of magnetosonic modes.
In this region gas orbits the star faster than the pattern of acoustic
modes, and every time a fluid element encounters a weak shock (into
which modes evolve) it loses angular momentum and moves towards 
the star, giving rise to continuing accretion.
Fig. \ref{mdotplots} shows plots of the instantaneous mass accretion
rate through radius $\cp$, $\dot{M}(\cp)$, for simulations
M9d,e,f (panels a,b,c) and for a $M=9$ hydro simulation from
\citet{BRS1} (panel d). The solid curves correspond to
periods when magnetosonic/acoustic modes are excited to a high
amplitude (high shock accretion rate)
and the dashed curves to periods when their amplitude is not as
high (low shock accretion rate). Each of the solid curves in panels
a-c corresponds
to one of the gap formation or opening events in Fig. \ref{densspacetime}.

In panels a-c, the dashed and solid curves converge to approximately the
same value of $\dot{M}$ in the outer disk in Fig. \ref{mdotplots}, but
the solid curves have a
significantly higher amplitude than the dashed ones in the inner disk ($\cp
\gtrsim 1$). The extra contribution to $\dot{M}$ for the
solid curves in the inner disk in panels a-c is provided by
magnetosonic modes, and
the base level of $\dot{M}$ in the outer disk is due to MRI
turbulence. 

Evidence for this assertion is provided by considering
panel d of Fig. \ref{mdotplots}. Since the simulation corresponding
to panel d is hydrodynamical, there is no MRI in the disk. This
explains why the dashed line is everywhere approximately zero, since
in the absence of waves excited to a high amplitude or MRI, there is no
efficient mechanism for angular momentum transport in the disk. On the
other hand, the shape and amplitude of the solid curve in the inner
disk in panel d are similar to those of the solid curves in panels
a-c. This shows that in the MHD case, the mass accretion rate in
the inner disk due to shocks can dominate that due to MRI, during
periods when modes are excited to a high amplitude.

Indeed, we can write the following expression for 
the combined mass accretion rate close to the star (interior to the
evanescent region in the disk) driven by both the MRI and shocks: 
\ba
\dot M=\frac{\Sigma s^2}{\Omega}
\left[4\pi\alpha q\frac{\partial\ln\left(\alpha q s^2\Sigma\varpi^2\right)}
{\partial\ln\varpi}+\frac{m}{3(2-q)}\left(1 + \frac{\Delta
  \Sigma}{\Sigma} \right)^{-2}
\left(\frac{\Delta\Sigma}{\Sigma}\right)^3\right], 
\label{eq:dotM}
\ea
where $q(\varpi)\equiv -d\ln\Omega/d\ln\varpi$ is the local 
shear rate ($q=3/2$ for a Keplerian disk but is a function of 
$\varpi$ for a general rotation profile),
$m$ is the azimuthal wavenumber of the mode, and $\Delta\Sigma/\Sigma$
is the density contrast across the shock (mode) fronts. The first
term in brackets is the contribution from MRI ($\alpha$ here is solely 
due to MRI) and can derived using the results of \citet{LyndenBellPringle}
and \citet{Rafikov2012}. The second term is the contribution from 
magnetosonic waves following from \citet{BRS}, namely from their 
equation (B7). For a characteristic azimuthal
wavenumber of $m\sim 15-30$ and MRI viscosity $\alpha\lesssim 10^{-2}$
(see Fig. \ref{alphaBspacetime}) measured in our simulations, one 
finds that both transport mechanisms provide comparable 
contribution to $\dot M$ if $\Delta\Sigma/\Sigma\sim 0.2$.

However, during periods of strong wave excitation $\Delta\Sigma/\Sigma$ 
increases and the wave term strongly dominates $\dot M$ because of 
the steep (cubic) dependence of $\dot M$ upon
$\Delta\Sigma/\Sigma$. For instance, the
amplitude of the $m=14$ magnetosonic mode changes by a factor of $\approx 5$ in
simulation M9f between $t=900$ and $t=1100$. This results in a
predicted mass
accretion rate due to shocks in the inner disk that is higher by
approximately two orders of magnitude at $t=1100$ as compared to
$t=900$. 

This effect of enhanced mass accretion rate by shocks during periods
when magnetosonic modes are excited to a high amplitude is
demonstrated by the solid curves in Fig. \ref{mdotplots}.
Equation (\ref{eq:dotM}) also explains the characteristic spatial 
dependence of $\dot M$ on $\varpi$: approximate conservation of the 
wave angular momentum flux (neglecting for the moment dissipation at
the shock fronts) results in (see equation (25) of \citet{BRS})
\ba
\frac{\Delta\Sigma}{\Sigma}\propto \left[\frac{|\Omega(\varpi)-\Omega_P|}
{\Sigma s^3\varpi}\right]^{1/2}.
\label{eq:contrast}
\ea
This expression shows that $\Delta\Sigma/\Sigma$ is highest near 
the BL, where $|\Omega(\varpi)-\Omega_P|$ is large and goes to zero 
in the evanescent region where $\Omega=\Omega_P$. This explains why
during the high wave amplitude episodes, when mass accretion
is clearly dominated by dissipation of acoustic modes, $\dot M$ is 
largest near the BL, but rapidly decays to small value as $\varpi$ 
approaches the evanescent region in the disk. 

Note that inside the BL MRI does not operate, $\alpha=0$, and 
the first term in equation (\ref{eq:dotM}) vanishes, leaving wave 
dissipation as the only means of mass transport. On the other hand, 
in the disk outside the corotation radius, waves do not contribute 
to transport and second term in equation (\ref{eq:dotM}) disappears.
Thus, MRI remains the only mass transport mechanism outside the 
resonant cavity in which the pattern of standing acoustic modes 
exists, in agreement with the picture outlined in \citet{BRS}
and \citet{BRS1}.


\subsection{Angular Momentum Transport due to Waves and MRI}
\label{angmomsec6}

In this section, we connect the density gap in the inner disk to
magnetosonic modes and discuss angular momentum transport in the star,
BL, and inner disk more generally. 

\begin{figure}[!h]
\centering
\subfigure{\includegraphics[width=0.8\textwidth]{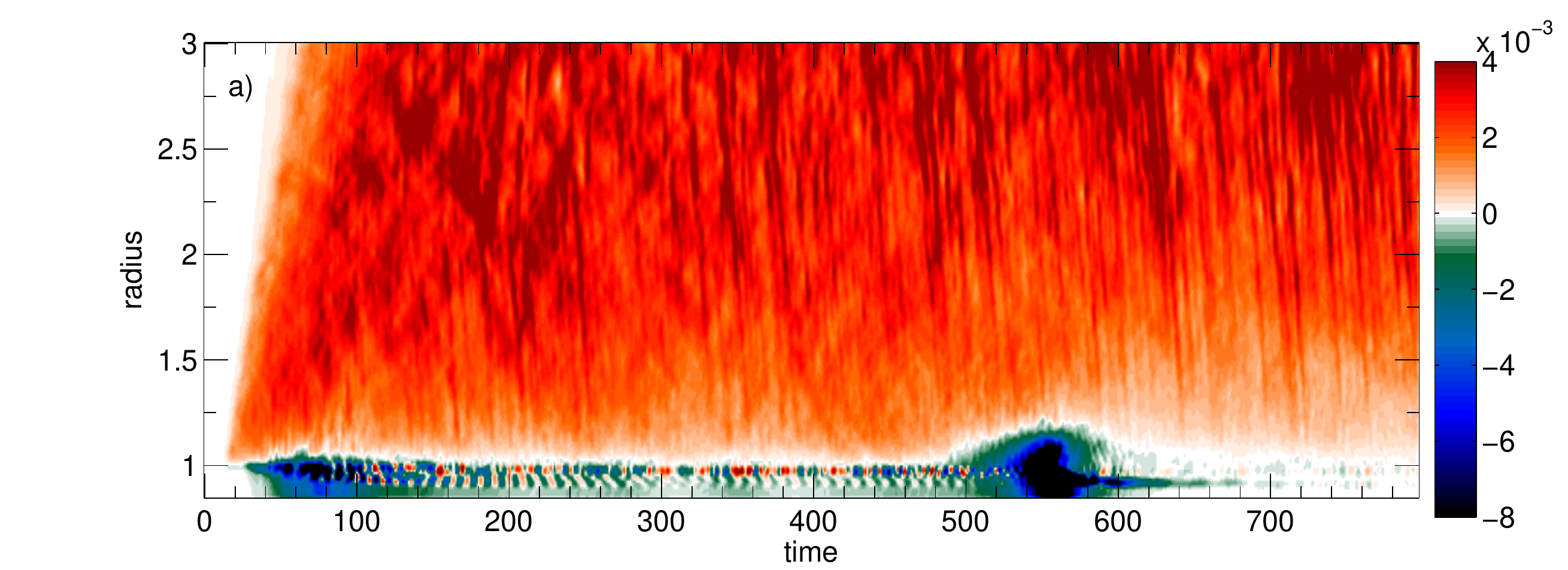}}
\subfigure{\includegraphics[width=0.8\textwidth]{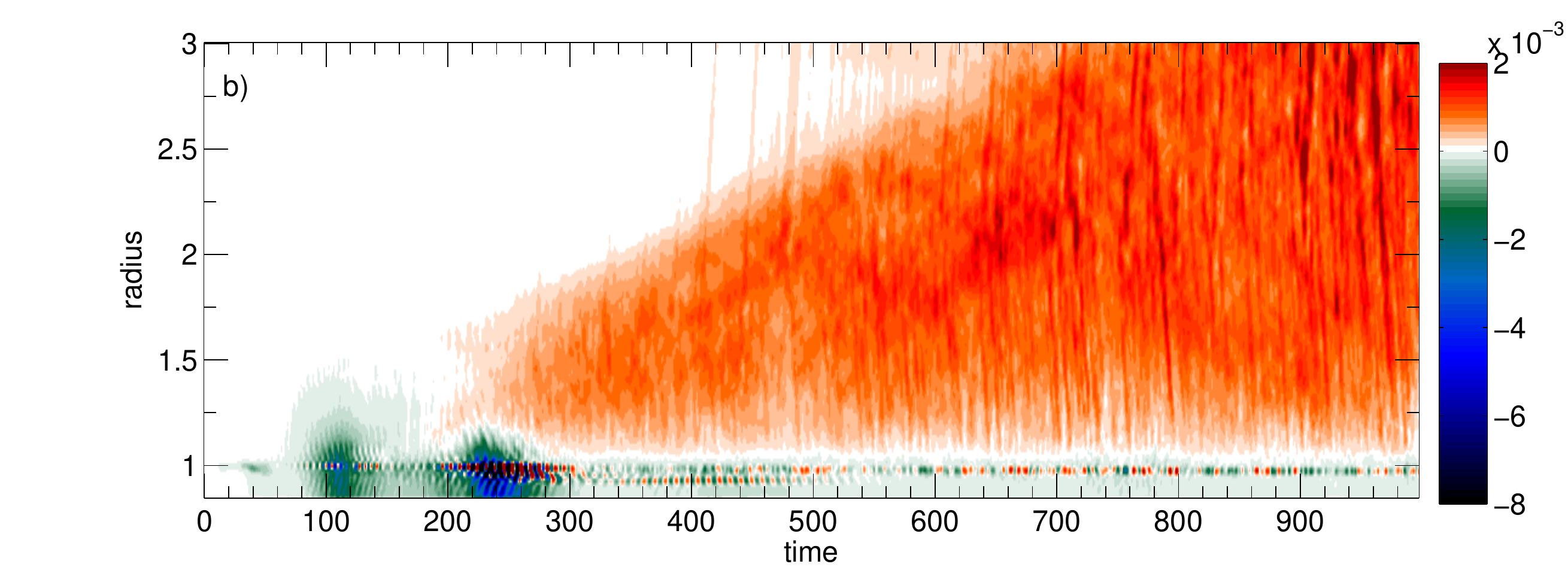}}
\subfigure{\includegraphics[width=0.8\textwidth]{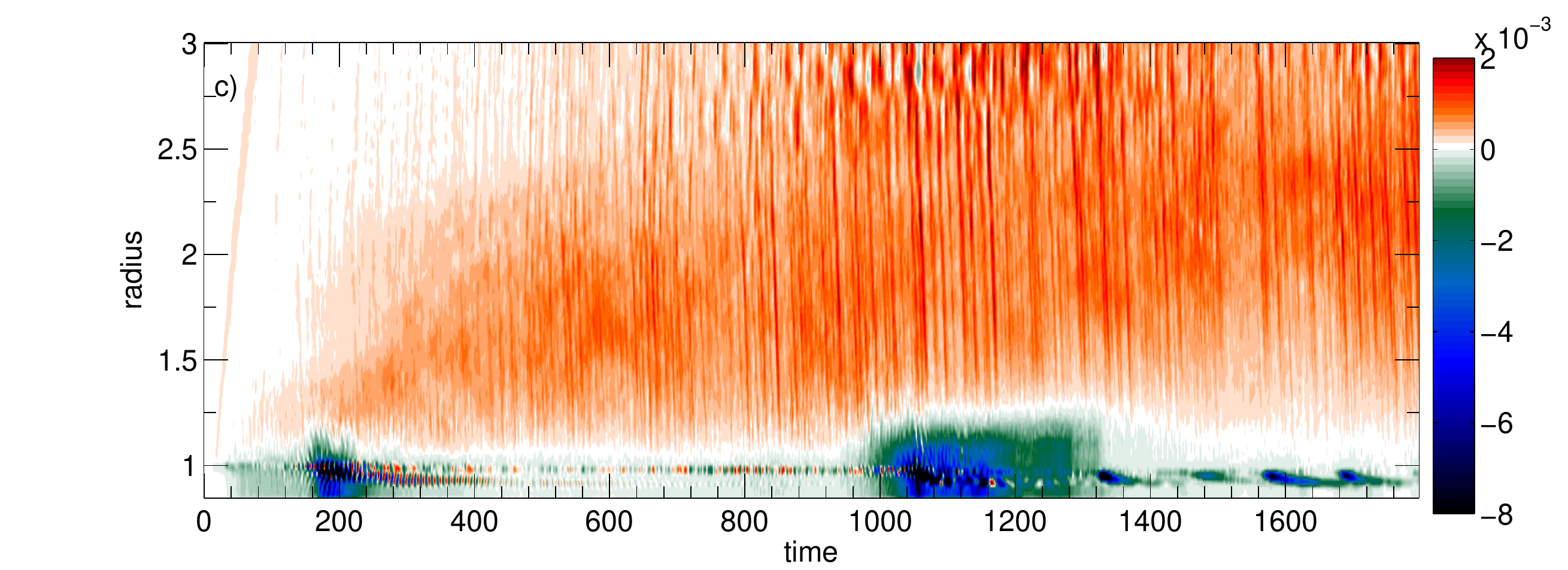}}
\caption{Panels a,b,c show radius time plots of the angular momentum
  current $C_S$ from simulations M9d,e,f (NVF, NAF, ZNF)
  respectively. Notice that
  $C_S$ changes sign in the vicinity of the BL with $C_S < 0$ for $\cp
  \lesssim 1$ and $C_S > 0$ for $\cp \gtrsim 1$.}
\label{stressspacetime6}
\end{figure}

Fig. \ref{stressspacetime6} shows
radius time plots of the angular momentum current, $C_S$, from
simulations M9d,e,f. In the disk, far away from the BL, angular
momentum transport is due primarily to
MRI turbulence (\S \ref{disksec}), and $C_S$ is approximately constant at a
given radius once the MRI has saturated. We note that the striations
appearing in $C_S$ are caused by waves. The striations in the panels of
Fig. \ref{stressspacetime6} are predominantly downward sloping in the
inner disk, indicating inward propagating waves. These waves are excited
at the ``front'' of MRI turbulence which propagates radially outward through
the disk over the course of the simulation. Although potentially a
nuisance, these waves do not play a major role in angular momentum
transport in our simulations.

In contrast to the outer disk, where the value of $C_S$ at a given
radius is
approximately constant in time once MRI turbulence has saturated,
$C_S$ varies by more than an order of magnitude in the star and inner
disk over the course of a simulation. This implies a mechanism of
angular momentum transport
that is decoupled from MRI turbulence. As we shall shortly
demonstrate, this mechanism is transport of
angular momentum by magnetosonic modes. Given this fact, we can
directly link magnetosonic modes to the gap formation and deepening
events in Fig. \ref{densspacetime}, since these are
simultaneous with the periods when $C_S$ is large in the BL and star
in Fig. \ref{stressspacetime6}.

\begin{figure}[!h]
\centering
\subfigure[]{\includegraphics[width=0.49\textwidth]{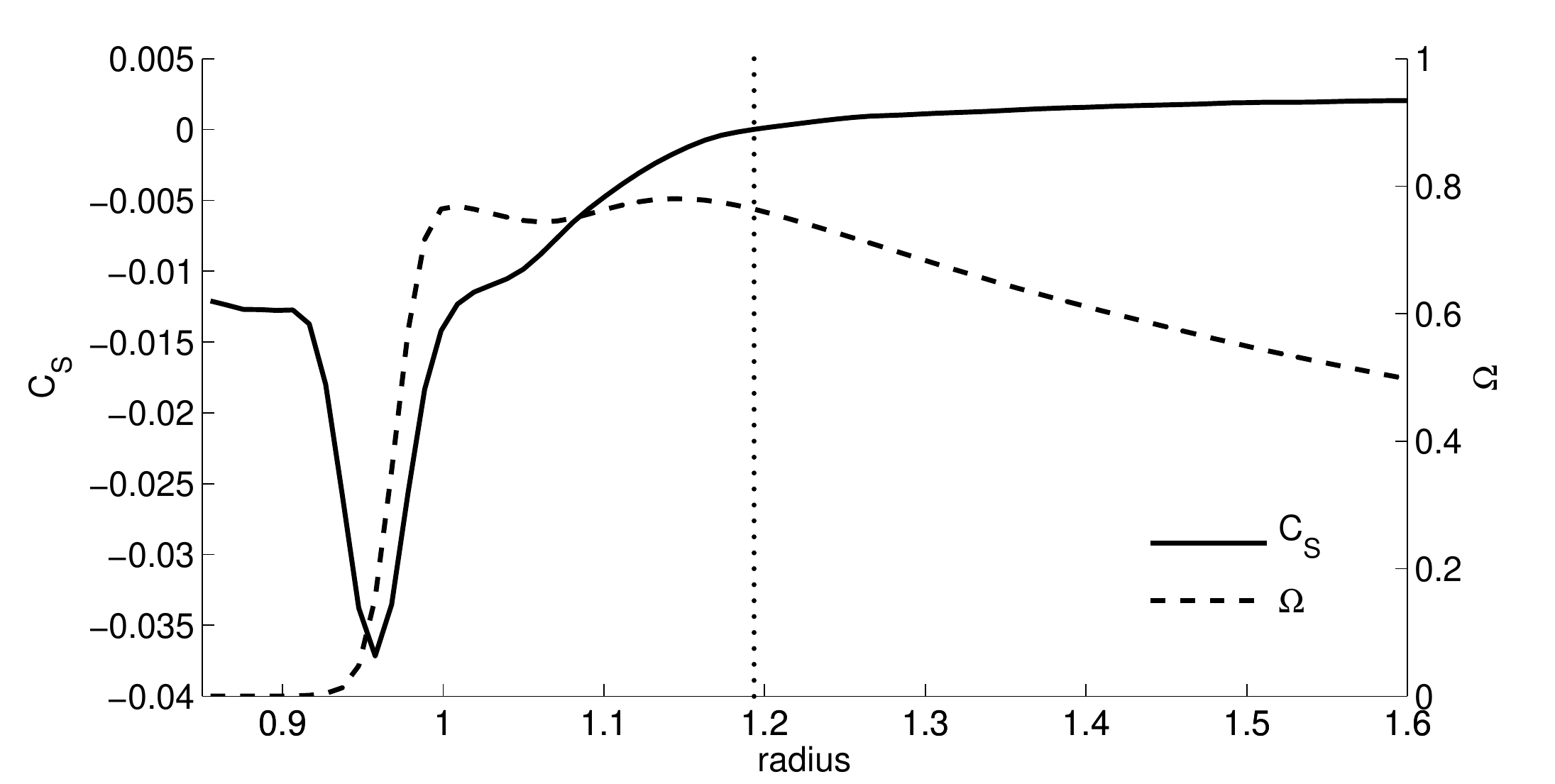}}
\subfigure[]{\includegraphics[width=0.49\textwidth]{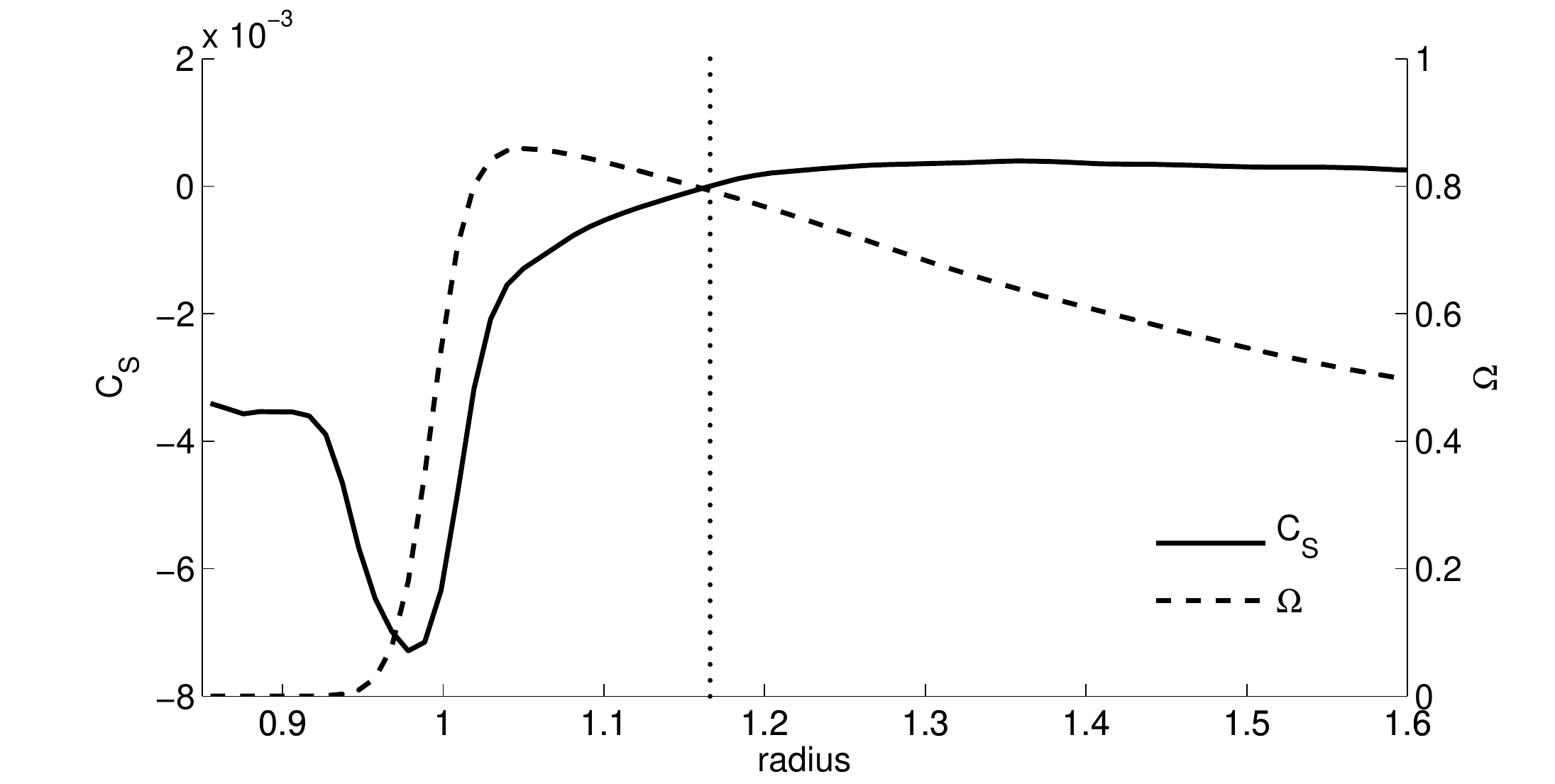}}
\subfigure[]{\includegraphics[width=0.49\textwidth]{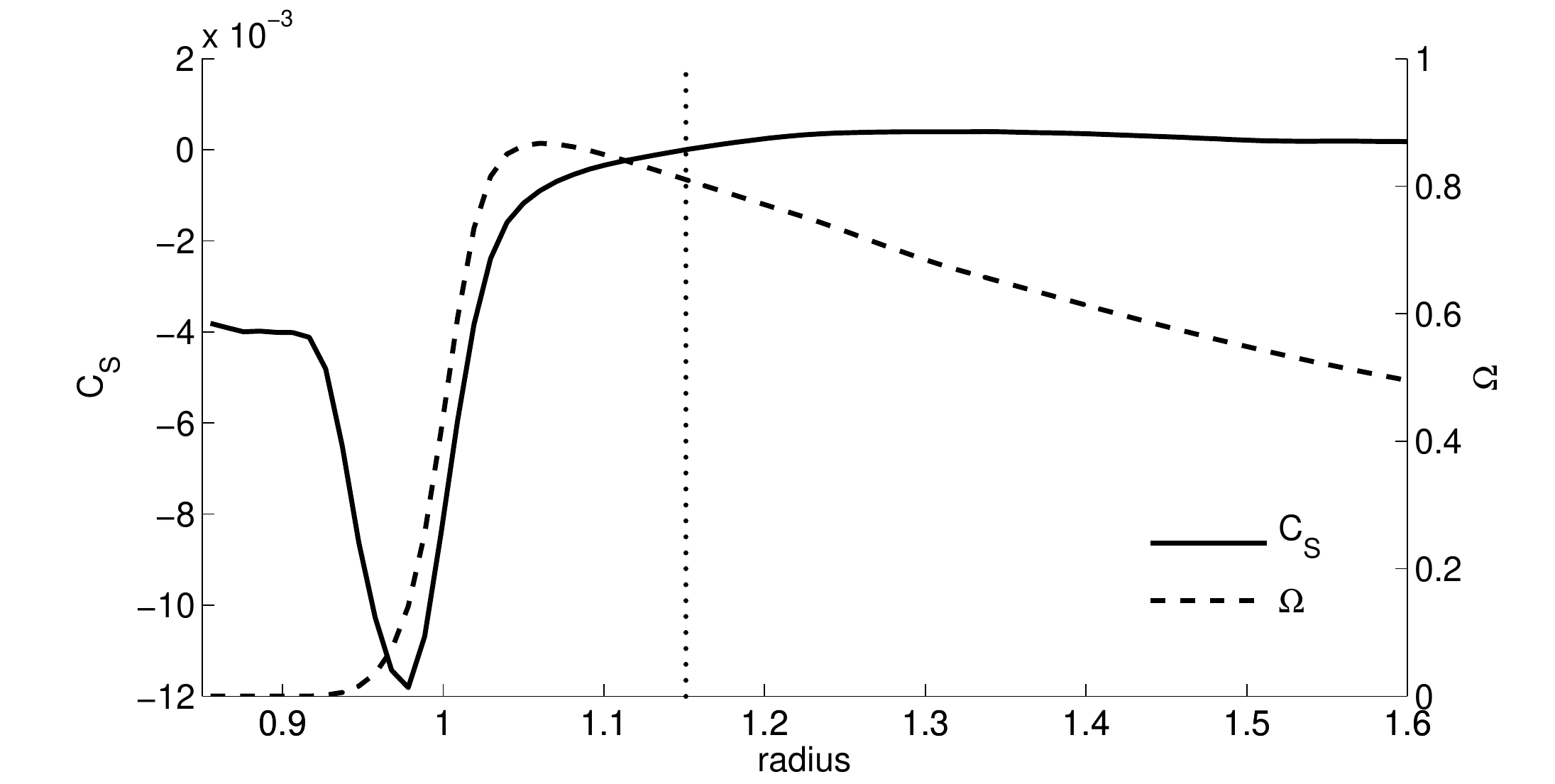}}
\caption{Panels a,b,c show plots of $C_S$ and $\Omega$ for simulations
M9d,e,f at times $t = 550$, 250, and 1100. The dashed vertical line denotes the
radius at which $C_S = 0$ and is significantly displaced into the disk
relative to the radius at which $d \Omega/d \cp = 0$. In general,
when angular momentum in the disk is transported by both a
magnetosonic mode and turbulent viscosity, 
the location at which $C_S = 0$ occurs between the radius at which $d
\Omega/d \cp = 0$ and the corotation radius of the mode,
$\cp_\text{cr} \approx 1.8$ in this example.}
\label{csomega}
\end{figure}

In addition to waves dominating angular momentum transport in
the star and BL, we find that they also contribute significantly
to angular momentum transport in the inner
disk. Fig. \ref{csomega} shows $C_S$ and $\Omega$ superimposed on the
same plot for simulations M9d,e,f
at times $t=550$, 250, and 1100, when the angular momentum current due to waves
is high in each simulation. The dashed vertical
line denotes the radius at which $C_S = 0$, and for an anomalous viscosity 
model, $C_S = 0$ and $d \Omega/d \cp = 0$ occur at the same radius (equation
[\ref{alphaSSeq}]). However, we see in Fig. \ref{csomega} that the
radius at which $C_S =
0$ is significantly displaced into the disk compared to the radius at
which $d \Omega/d \cp
= 0$. This implies that $C_S$ contains a contribution due to
waves in the disk, since $C_S < 0$ for outgoing waves up to the
corotation radius in the disk, which is located at
$\cp_\text{cr} \approx 1.8$ ($m=14$ mode) in each of the panels in
Fig. \ref{csomega}.

We conclude that
when waves are excited to a high amplitude, they significantly modify
the profile of $C_S$ in the inner disk compared to what is expected
for a
turbulent viscosity model. In particular, if both magnetosonic modes
and anomalous viscosity contribute appreciably to $C_S$, then the
radius at which $C_S = 0$ in the disk is neither
at the corotation radius of the mode, nor at the radius where $d
\Omega/d\cp = 0$, but somewhere in between.

We also mention that the variation of $C_S$ with radius in the star and
BL in each of the panels in Fig. \ref{csomega} is
consistent with $C_S(\cp)$ for the lower branch (Fig. 12 of
\citet{BRS1}). Since the lower branch is directly observed in our simulations
(\S \ref{acoussec}), this provides further confirmation that
angular momentum is transported by waves in the BL and star.

\subsection{Hydrodynamical Nature of Angular Momentum Transport in the Star
  and BL}

Having established that magnetosonic modes transport
angular momentum in the BL and star, and also influence transport
in the inner disk, we now demonstrate that the mechanism of angular transport
by acoustic modes is essentially hydrodynamical in nature. This
provides evidence for the argument presented in \S \ref{disrelsec} that 
when $\beta \gg 1$, the properties of acoustic modes are not
significantly affected by the magnetic field, and the results of
\citet{BRS1} are valid.

\begin{figure}[!h]
\centering
\subfigure{\includegraphics[width=0.49\textwidth]{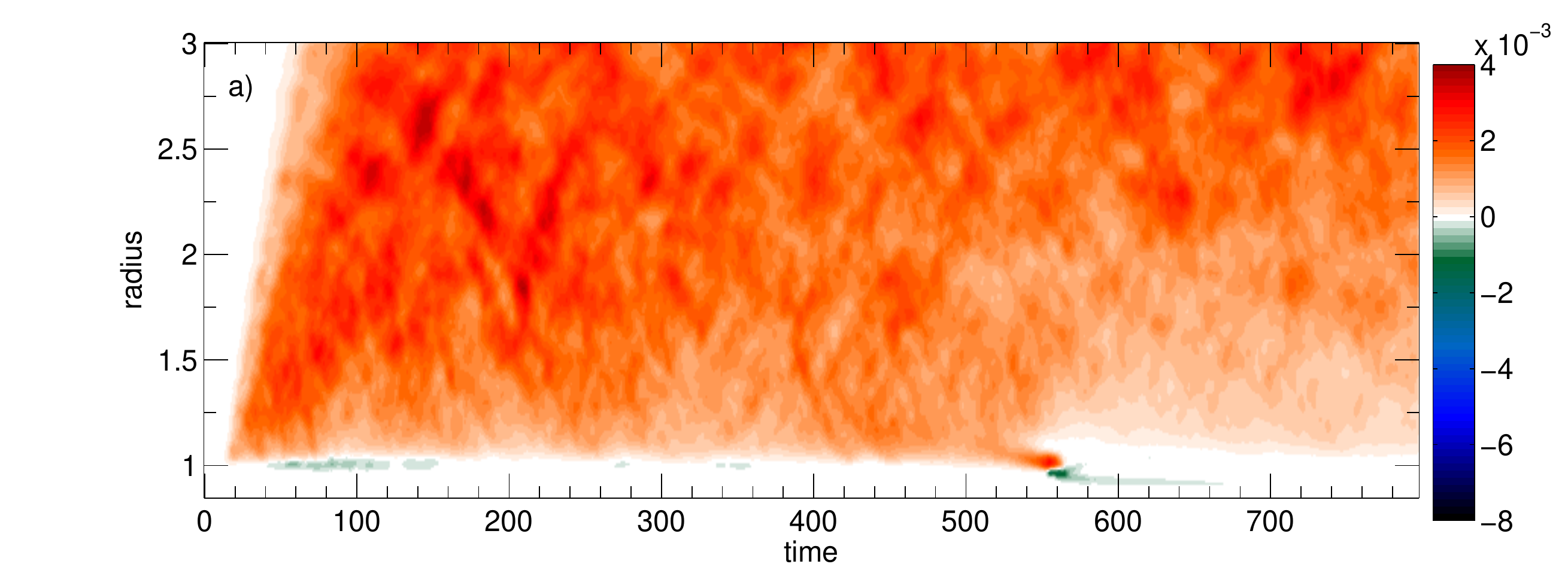}}
\subfigure{\includegraphics[width=0.49\textwidth]{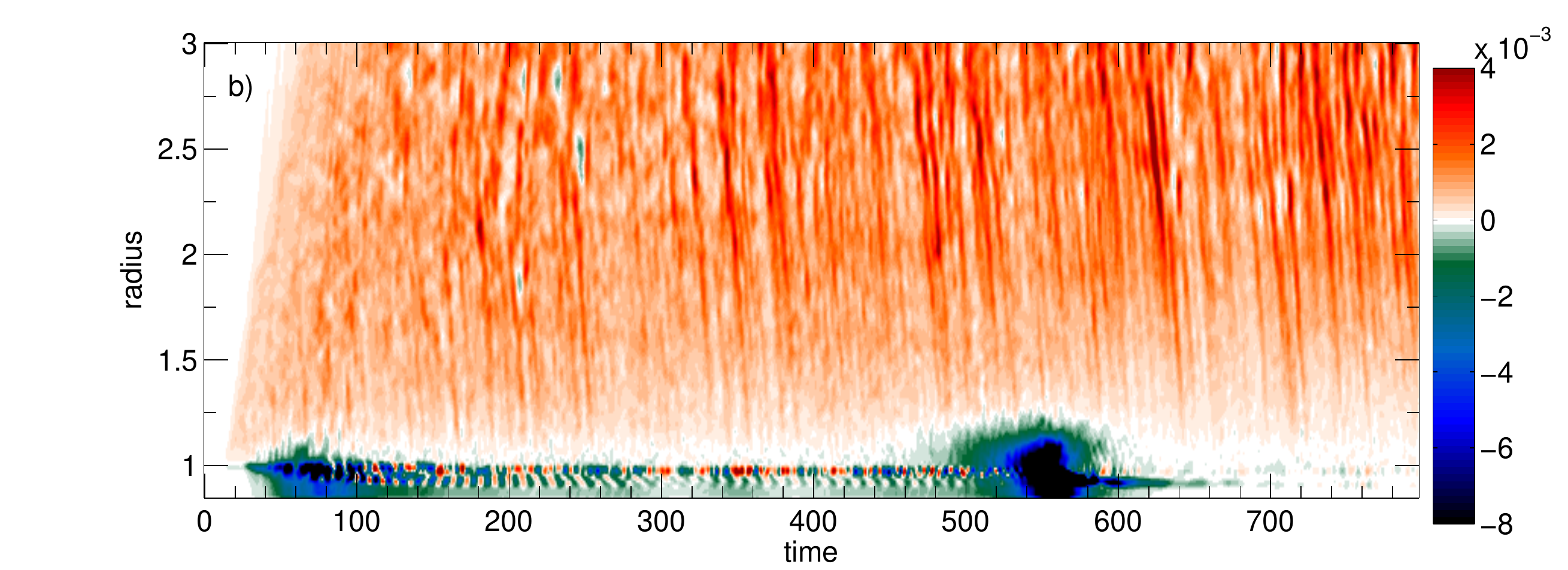}}
\subfigure{\includegraphics[width=0.49\textwidth]{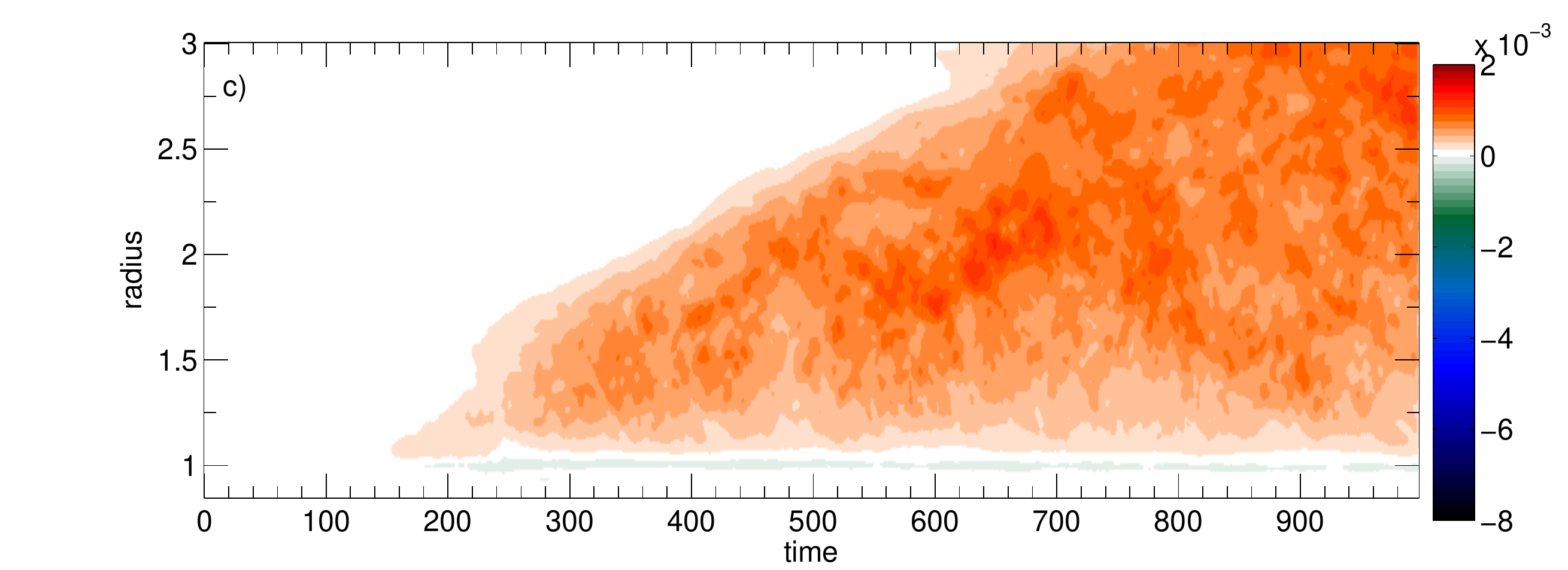}}
\subfigure{\includegraphics[width=0.49\textwidth]{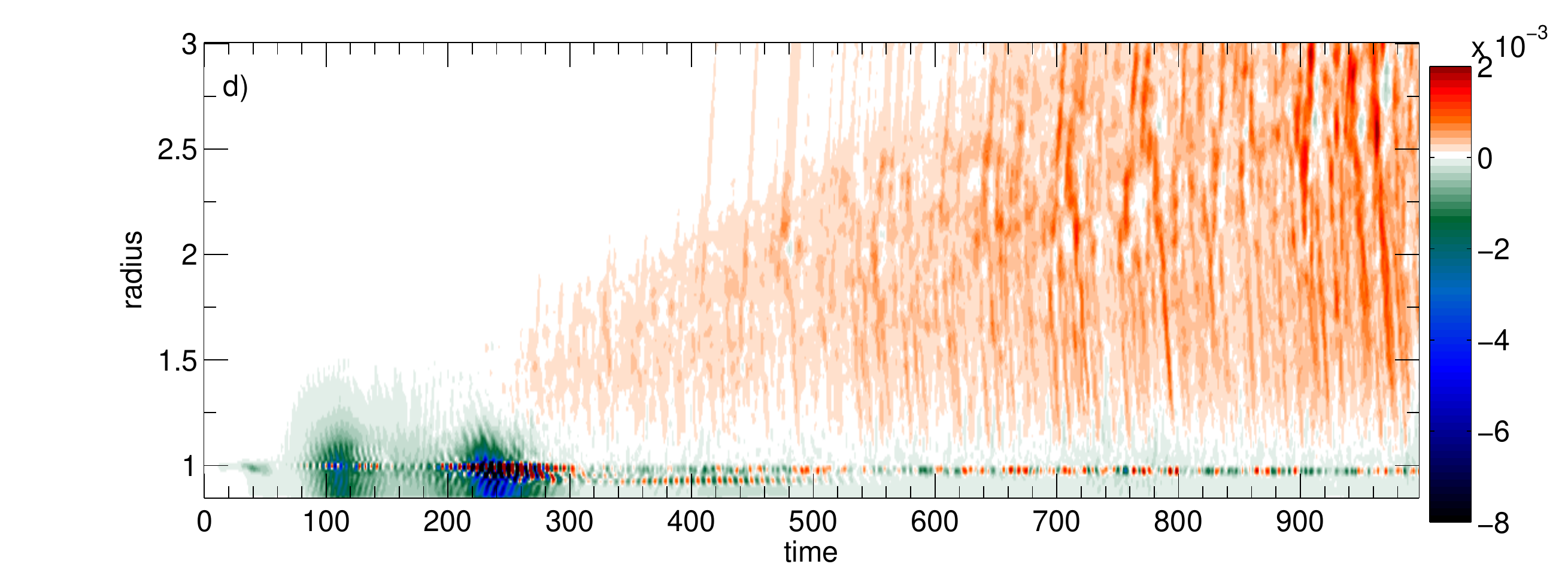}}
\subfigure{\includegraphics[width=0.49\textwidth]{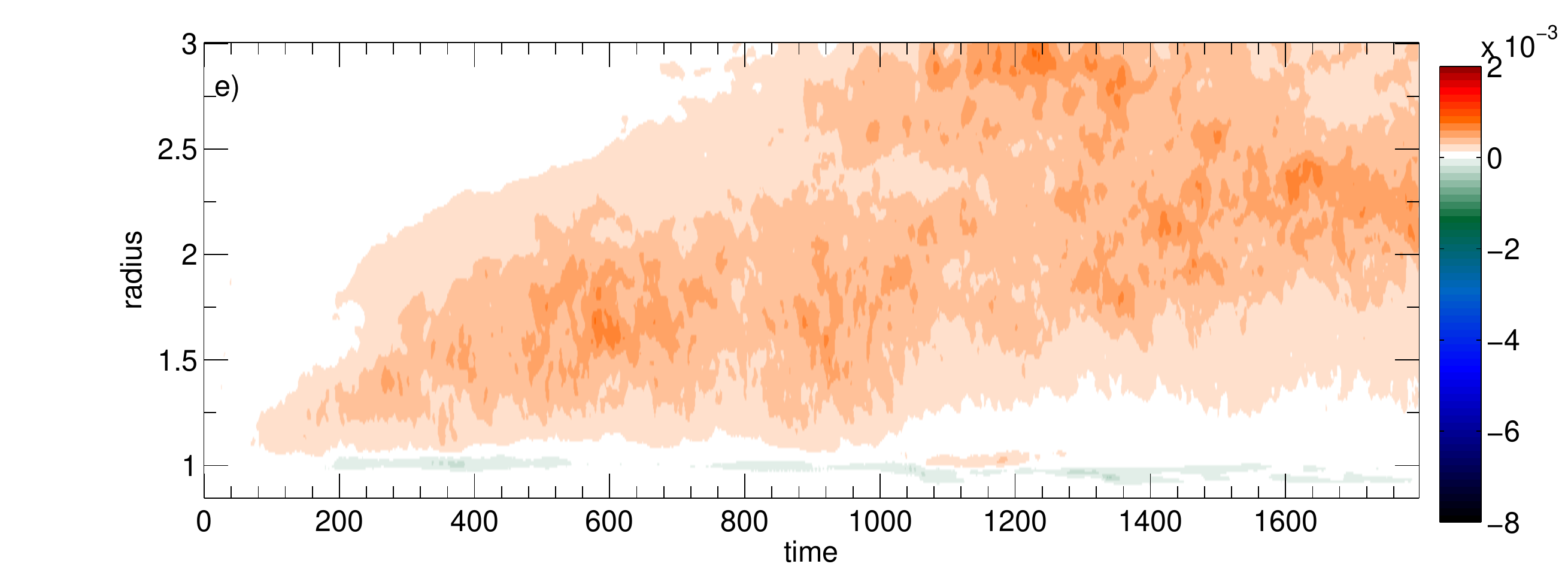}}
\subfigure{\includegraphics[width=0.49\textwidth]{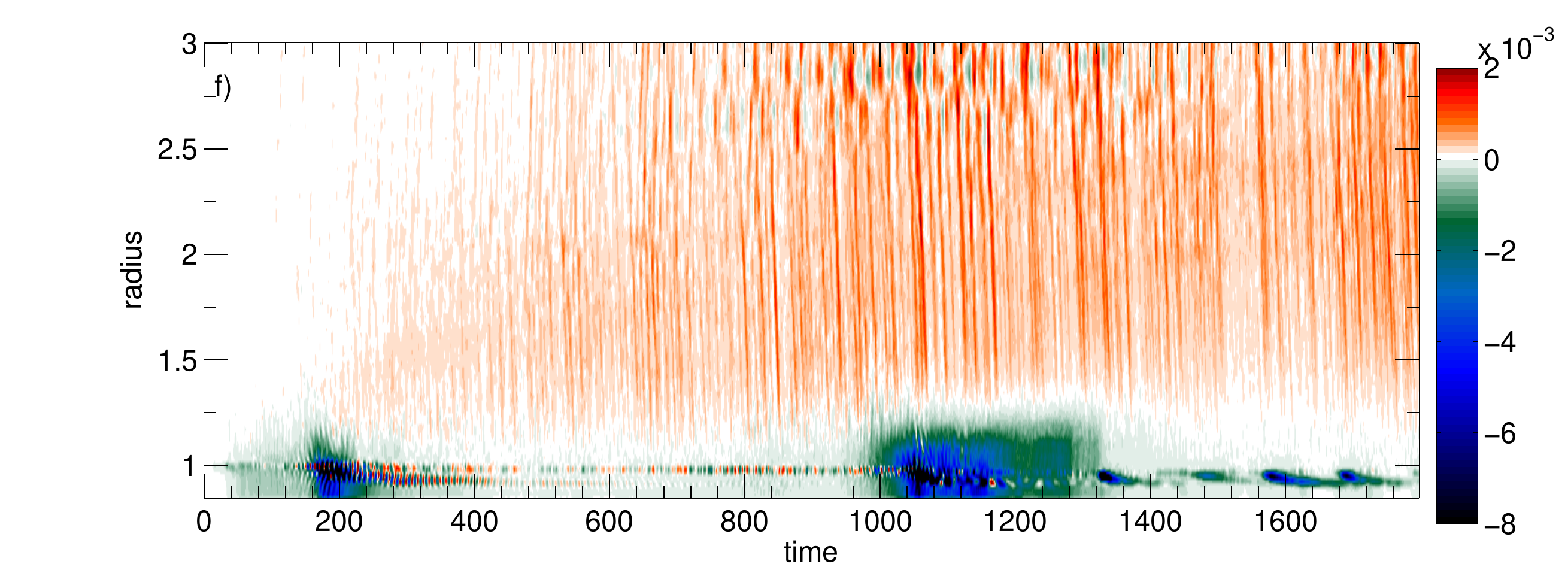}}
\caption{Panels a,c,e show radius time plots of $C_{S,B}$ from
  simulations M9d,e,f (NVF, NAF, ZNF) respectively, and panels b,d,f
  show radius time
  plots of $C_{S,H}$ from simulations M9d,e,f. The scale, colormap,
  and axes labels for each simulation (M9d,e,f) are the same as
  in Fig. \ref{stressspacetime6}. Notice that $C_{S,B}$
  vanishes inside the star, and there is no bump in $C_{S,B}$ within
  the BL.
  On the other hand, $C_{S,H}$, is large inside the BL and is both large and
  approximately constant (spatially) within the star, which suggests
  transport by waves.} 
\label{stressspacetimesplit}
\end{figure}

To show that stresses in the BL are hydrodynamical, we split
$C_S$ into magnetic and nonmagnetic
components (equation [\ref{CSsplit}]). Fig. \ref{stressspacetimesplit} shows
$C_{S,B}$ and
$C_{S,H}$ for each of simulations M9d,e,f, and several things are
immediately apparent. The
first is that $C_{S,B}$ is several times larger than $C_{S,H}$ in
the disk, which is consistent with the fact that the magnetic stress
is several times the hydrodynamical stress for MRI turbulence
\citep{Sorathia}. The second is that $C_{S,B}$ vanishes going
into the star and
there is no bump in $C_{S,B}$ within the BL. This means that magnetic
stresses are {\it not} important for transport in the BL. This is
consistent with the results of \citet{Pessah}, who
found that in the shearing sheet approximation, the magnetic stress
oscillates about zero for a rising rotation profile ($d \Omega/ d\cp >
0$), as in the BL. The third is that $|C_{S,H}| \gg |C_{S,B}|$ in
the star and BL, which is definitive proof that a hydrodynamical
mechanism of angular momentum transport operates there -- namely acoustic modes
weakly modified by a magnetic field. 

\subsection{Boundary Layer Width}
\label{BLwidthsec6}

\begin{figure}[!h]
\centering
\includegraphics[width=0.7\textwidth]{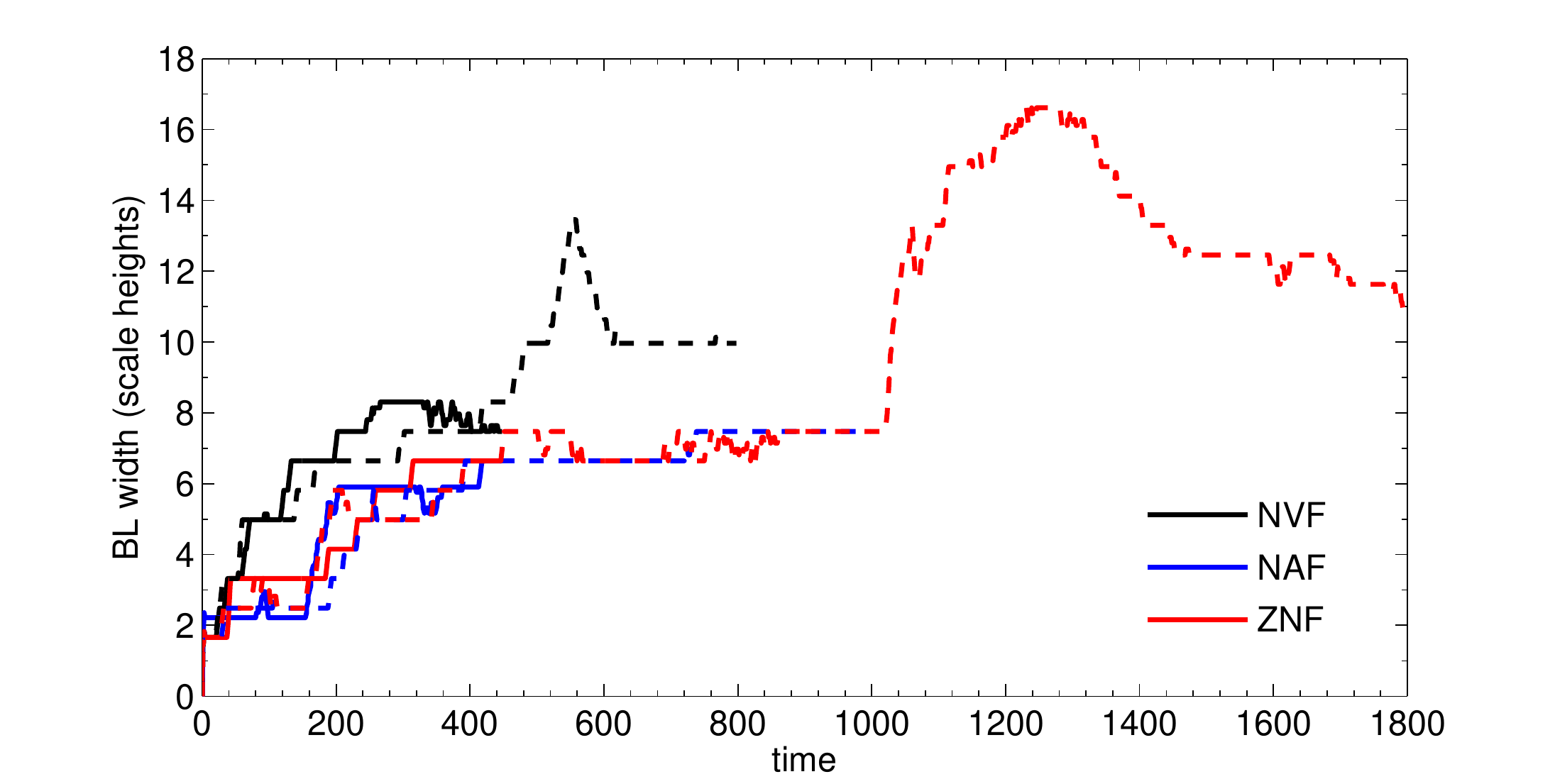}
\caption{Boundary layer width in units of scale height $s^{-2}$ 
(defined implicitly in equation
  [\ref{BLwidth6}]) as a function of
  time, for each of our simulations. The black, blue, and red curves
  correspond to NVF, NAF, and ZNF simulations respectively. The solid
  lines correspond to simulations with the higher resolution for a
  given magnetic field geometry.}
\label{blwidthfig6}
\end{figure}

\citet{BRS} and \citet{BRS1} found that in hydro simulations the
BL settled to a radial width of $6-7$ radial scale heights
($h_s \equiv s^2/g(\cp_\star)$). Using their definition for the BL
as the region in which
\ba
\label{BLwidth6}
0.1 < \langle v_\phi(\cp)\rangle/v_K(\cp) < 0.9, 
\ea
we confirm that after the initial gap opening event, the BL in our MHD
simulations also
settles down to a radial width of 6-7 radial scale heights
(Fig. [\ref{blwidthfig6}]). However, the BL undergoes
widening events that are coincident with gap deepening events (\S
\ref{gapsec}) and the periods when $C_S$ is large in the star and BL (\S
\ref{angmomsec6}). Stochastic widening events were also observed by
\citet{BR} in 2D hydro simulations, and it appears that in both hydro
and MHD cases these events are mediated by the action of shocks on the
fluid in the inner disk.

The fact that the BL widening events occur simultaneously with gap
deepening events, suggests a common origin, implying that magnetosonic
modes play a critical role in regulating the width of the BL. However,
given the occurrence of stochastic widening events in the simulations,
the width of the BL is time-variable and it is not possible to assign
a single number to it. Nevertheless, what the simulations suggest is
that the BL initially settles down to a steady state width of 6-7
radial scale heights and is subsequently widened, during periods when
magnetosonic modes are excited to a high amplitude.

\label{blwidth}

\section{Discussion}
\label{discussion}

We have shown that even in the presence of MRI turbulence in the disk,
waves dominate angular momentum transport in the
BL, and influence transport in the inner disk. The importance of waves
for angular momentum transport in the BL was
previously shown by \citet{BRS1} for the
purely hydrodynamical case. 

We find that when a magnetic field is
introduced, the acoustic modes discussed by \citet{BRS1} become
magnetosonic modes. These magnetosonic modes have a dispersion
relation that differs from the dispersion relation for acoustic modes
by factors that are
$\mathcal{O}(\beta^{-1})$. Since the field in our simulations is
highly subthermal ($\beta^{-1} \lesssim .06$) everywhere, including the BL, the
hydrodynamical theory of
\citet{BRS1} remains a good approximation to the dynamics of
magnetosonic modes in our simulations. 

Unlike MRI turbulence, which can be parametrized by
an anomalous viscosity, the transport of energy and angular 
momentum in the BL is a {\it non-local} process, as has been emphasized 
in \citet{BRS1}. Energy and momentum are carried away from the BL by
acoustic waves weakly modified by magnetic field 
into both the star and the disk and can 
be deposited far away from the BL. This has important implications for 
stellar spinup, heating of the star, disk and the BL itself, and
for observational manifestations of the BL phenomenon \citep{BRS1}. 
In the MHD case, additional dissipation is produced in the disk 
due to the operation of MRI but, as one can infer from Fig. 
\ref{alphaBspacetime} where $\alpha_B$ is seen to vanish in the BL, 
the MRI-generated heat is unlikely to penetrate into the BL and 
affect its structure.

Amplification of the magnetic field in the BL that we see in our
simulations (\S \ref{sect:Bampl}) is consistent with earlier 
theoretical expectations \citep{Pringle1989} but does not result 
in efficient 
magnetically-mediated transport of angular momentum in this region. 
Our study is in agreement with the analytical 
work of \citet{Pessah}, who analyzed the evolution of MHD modes
in presence of shear with $\partial\Omega/\partial\varpi>0$
and demonstrated that although the energy density of these modes 
can be amplified significantly, their associated stresses oscillate 
around zero, rendering them inefficient at transporting 
angular momentum. As a consequence,
angular momentum transport in the BL is mediated by quasi-acoustic 
modes, and in particular shock dissipation of these modes in the
inner disk.

Previously, \citet{SteinackerPapaloizou}, \citet{Armitage} performed
isothermal MHD simulations of the BL, and the setup of
\citet{Armitage} in particular was nearly identical to
ours. Therefore, a natural question that arises is why did these
authors not observe acoustic modes and recognize their importance for
angular momentum transport in the BL? Interestingly, the answer to this
question may be that the modes were in fact observed in these early
simulations, but their presence and significance were not fully appreciated.

For instance, Fig. 3 of \citet{Armitage} shows an image of relative 
density perturbation averaged over the z-dimension. The density
fluctuations in the image exhibit a leading/trailing mode pattern 
that looks strikingly similar to the lower mode trapped between the
BL and the evanescent region in the disk, see Fig. \ref{znf_snapshot}. 
Note that $\Delta \Sigma/\Sigma \propto\Sigma^{-1/2}$ for an 
acoustic mode propagating into the star by angular momentum 
conservation, see equation (\ref{eq:contrast}),  which explains 
why the amplitude of density fluctuations rapidly decreases going 
into the star in Fig. 3 of \citet{Armitage}. 

It is harder to comment on whether \citet{SteinackerPapaloizou}
observed acoustic modes, since they used a different initial setup
than ours and {\it prescribed} a rotation profile for the BL region
in their simulations. In contrast, the profile of $\Omega$ in the BL 
is naturally established in our simulations  due to action of the 
sonic instability early on in the simulations \citep{BRS,BRS1}, 
which widens the initial interfacial
region into a self-consistent BL. Nevertheless, it is quite possible
that \citet{SteinackerPapaloizou} also observed acoustic modes excited in
the BL, as they briefly mentioned the presence of ``stochastic spiral
patterns'' in their simulations, which could well have been the
manifestation of acoustic modes (e.g. see Fig. 7 of
\citet{SteinackerPapaloizou}).

Another important question that arises is given that we have used an
isothermal equation of state,
what might be expected for a more realistic equation of
state? As discussed in \citet{BRS1} one pathology of the isothermal
equation of state is that the Brunt-V\"ais\"al\"a frequency, $N^2 =
(\gamma-1)g^2/s^2$, is equal to zero. This means that our
computational model does not support gravity waves and in a real star
with nonzero Brunt-V\"ais\"al\"a frequency, magnetosonic waves in the
disk could couple to gravity waves in the star increasing the number
of possible modes in the star-BL-disk system.

Another difference is that the sound speed is not constant in a real
star and is an inverse function of the
radius. This means that the cutoff frequency below which sound waves
do not propagate is also an inverse function of the radius and without
a damping process, such as radiative damping, waves propagating into
the star would reflect back towards the BL. A simple way to model this effect
in simulations would be to introduce a reflecting wall at the inner
boundary of the simulation domain. \citet{Glatzel}, \citet{BR} found
analytically that for the case of a supersonic shear layer with a
linear velocity profile, instability was present for both
reflecting wall and radiation boundary conditions. Therefore, although
reflection at an inner boundary would likely modify the details of
angular momentum transport by waves, in particular the spinup rate of the star
and possibly the wavelength of the dominant mode,
magnetosonic modes would still be excited in such a system.

\acknowledgements

Resources supporting this work were provided by the NASA High-End 
Computing (HEC) Program through the NASA Advanced Supercomputing 
(NAS) Division at Ames Research Center. We thank Jeremy Goodman and 
Anatoly Spitkovsky for useful discussions. The financial support 
for this work is provided by NASA grant 
NNX08AH87G.


\appendix
\section{Dispersion Relation for the Compressible Vortex Sheet in the Presence
  of a Constant Magnetic Field that is Oriented Perpendicular to the Motion}
\label{disapp}

Here we present the dispersion relation for modes of the compressible
vortex sheet with wavevectors lying in the $x$-$y$ plane, when a
$z$-directed magnetic field is present. Under this set of assumptions, the
$\bfB \cdot
\bfnabla \bfB$ term in the MHD equations is zero, which
means Alfv\'{e}n waves are not permitted, the slow wave is null, and
only the fast
magnetosonic wave survives. In fact, in such a setup the magnetic
field simply provides
pressure support and behaves as though it were a fluid with an
effective adiabatic index of $\gamma_B = 2$. Although this is not
a realistic description for astrophysical boundary layers, it gives us a
flavor for
how the dispersion relation of acoustic modes is modified when a
weak magnetic field is introduced. In particular, it shows that for a
weak magnetic field, the corrections to the dispersion relation are small
and are of order $\mathcal{O}\left(\beta^{-1}\right)$.

Since all fluid quantities in our setup are independent of $z$ we are
essentially studying a 2D problem. In that case, the ideal MHD
equations with an adiabatic equation of state can be formulated as
\ba
\label{eq:cont1}
\frac{\partial \rho}{\partial t} + \bfnabla \cdot (\rho \bfv) &=& 0,
\\
\label{eq:mom1}
\frac{\partial (\rho \bfv)}{\partial t} + \bfnabla \cdot (\rho \bfv
\bfv) &=& -\bfnabla \left( P + \frac{B_z^2}{2\mu} \right), \\
\label{eq:state1}
P &=& K\rho^{\gamma}, \\
\label{eq:induction1}
B_z &=& \sigma \rho,
\ea
where, we have replaced the induction equation (\ref{eq:induction})
with the frozen-in law (\ref{eq:induction1}) and introduced the
``magnetization'' of the fluid per unit mass, $\sigma$. Equations
(\ref{eq:cont1}-\ref{eq:induction1}) can be simplified to
\ba
\label{eq:cont2}
\frac{\partial \rho}{\partial t} + \bfnabla \cdot (\rho \bfv) &=& 0,
\\
\label{eq:mom2}
\frac{\partial (\rho \bfv)}{\partial t} + \bfnabla \cdot (\rho \bfv
\bfv) &=& -\bfnabla \left( K \rho^\gamma + \frac{\sigma^2}{2\mu}
\rho^2 \right).
\ea
Written in this form, it is clear that for our setup the magnetic
field behaves in
the same manner as a fluid with $\gamma_B = 2$.

We now describe the compressible vortex sheet setup, following the
notation of \citet{BR}. The velocity profile of the model setup is
given by
\ba
\label{vsheet}
V_y(x) = \left\{
     \begin{array}{lr}
       \bar{V}_y, \ x > 0 \\
      -\bar{V}_y, \ x < 0,
     \end{array}
   \right.
\ea
where $\bar{V}_y$ is a constant, and a velocity discontinuity is
present at $x = 0$.
For $x >0$, the density, magnetization, and sound speed are constant
and are given by $\rho_+$, $\sigma_+$, and $s_+$,
respectively. Similarly, for $x<0$
these parameters are constant as well and are given by $\rho_-$,
$\sigma_-$, and $s_-$, respectively. 

Using the techniques laid down in \citep{Gerwin,Alexakis,BR}, it is a
straightforward if lengthy exercise to derive the dispersion relation
for linearized perturbations of this system. Thus, we omit the
derivation and simply state the dispersion relation, which is
\ba
\label{disrelB}
\eps^2\left[(M_\text{ms}+\varphi)^2
  \eps^{-1}(1+\delta)^{-1}-1\right]\left(M_\text{ms}-\varphi\right)^4 =
\left[(M_\text{ms}-\varphi)^2-1\right]\left(M_\text{ms}+\varphi\right)^4.
\ea
Here, $\eps \equiv \rho_+/\rho_-$ is the density contrast across the vortex
  sheet, $\varphi \equiv \omega/k_ys_+$ is the dimensionless phase speed,
$M_\text{ms} \equiv \bar{V}_y/\sqrt{s_+^2 + v_{A,+}^2}$ is the magnetosonic
  Mach number, and $\delta$
is defined through the relation
\ba
\eps(1+\delta) &=& \frac{s_-^2 + v_{A,-}^2}{s_+^2 + v_{A,+}^2}.
\ea
As in the text, $v_A$ denotes the Alfv\'{e}n velocity.

Equation (\ref{disrelB}) is identical to equation (39) of \citet{BR} for
the unmagnetized case, except that the Mach number
has been replaced by the magnetosonic Mach number,
and there is an additional factor of $(1+\delta)^{-1}$. The first
of these modifications, the replacement of the Mach number with the
magnetosonic Mach number, is to be expected and was argued for in the
text. The second is harder to anticipate and a nonzero $\delta$ can be
thought of as coming from a difference in the effective $\gamma$
of the fluid across the vortex sheet. The effective
$\gamma$ of the fluid is a suitably weighted combination of $\gamma$
and $\gamma_B$. 

The major point of equation (\ref{disrelB}) as regards astrophysical
BLs is to show that for a weak
magnetic field, the dispersion relation for the compressible vortex
sheet is only modified up to terms of
$\mathcal{O}\left(v_A^2/s^2\right)$ (i.e.
$\mathcal{O}\left(\beta^{-1}\right)$). Since the dispersion relation
for the three wave
branches of acoustic modes discussed in \S \ref{acoussec} is simply
related to the dispersion relation for 
the compressible vortex sheet \citep{BRS}, we reason that properties
of  acoustic modes are also only modified by terms of
$\mathcal{O}\left(\beta^{-1}\right)$ in the presence of a magnetic
field. Since $\beta$  tends to be large in our simulations, even in
the BL, we reason that
the magnetic field due to MRI perturbs the acoustic modes, but does
not modify them in a fundamental way.


\begin{thebibliography}

\bibitem[Alexakis et al.(2002)]{Alexakis} Alexakis, A., Young, 
Y., \& Rosner, R.\ 2002, \pre, 65, 026313 

\bibitem[Armitage(2002)]{Armitage} Armitage, P.~J.\ 2002, 
\mnras, 330, 895

\bibitem[Balbus \& Hawley(1991)]{BalbusMRI} Balbus, S.~A., \& Hawley,
  J.~F.\ 1991, \apj, 376, 214

\bibitem[Balbus 
\& Hawley(1992)]{Balbus92} Balbus, S.~A., \& Hawley, J.~F.\ 1992,
  \apj, 400, 610

\bibitem[Balbus 
\& Papaloizou(1999)]{BalbusPapaloizou} Balbus, S.~A., \& Papaloizou,
  J.~C.~B.\ 1999, \apj, 521, 650

\bibitem[Belyaev \& Rafikov(2012)]{BR} Belyaev, M.~A., \& Rafikov,
  R.~R. \ 2012, \apj, 752 115

\bibitem[Belyaev et al.(2012a)]{BRS} Belyaev, M.~A., Rafikov, R.~R., \&
  Stone, J.~M.\ 2012, \apj, 760, 22 

\bibitem[Belyaev et al.(2012b)]{BRS1} Belyaev, M.~A., 
Rafikov, R.~R., \& Stone, J.~M.\ 2012, arXiv:1212.0580 

\bibitem[Chandrasekhar(1960)]{Chandra}
Chandrasekhar, S. 1960, Proc. Natl. Acad. Sci., 46, 253

\bibitem[Coleman et al.(1995)]{CKK} Coleman, C.~S., Kley, 
W., \& Kumar, S.\ 1995, \mnras, 274, 171 

\bibitem[Fromang 
\& Papaloizou(2007)]{Fromang} Fromang, S., \& Papaloizou, J.\ 2007,
  \aap, 476, 1113

 \bibitem[Fromang et 
al.(2007)]{Fromang2} Fromang, S., Papaloizou, J., Lesur, G., \& Heinemann, T.\ 2007, \aap, 476, 1123

\bibitem[Fu 
\& Lai(2012)]{FuLai2012} Fu, W., \& Lai, D.\ 2012, arXiv:1212.2215 

\bibitem[Gerwin(1968)]{Gerwin} Gerwin, R.~A.\ 1968, Reviews of 
Modern Physics, 40, 652 

\bibitem[Ghosh 
\& Lamb(1978)]{GhoshLamb} Ghosh, P., \& Lamb, F.~K.\ 1978, \apjl, 223, L83

\bibitem[Glatzel(1988)]{Glatzel} 
Glatzel, W.\ 1988, \mnras, 231, 795

\bibitem[Goodman
\& Rafikov(2001)]{GoodmanRafikov} Goodman, J., \& Rafikov, R.~R.\
  2001, \apj, 552, 793

\bibitem[Guan et al.(2009)]{GuanGammie} Guan, X., Gammie, C.~F., 
Simon, J.~B., \& Johnson, B.~M.\ 2009, \apj, 694, 1010 

\bibitem[Hawley et al.(1995)]{HGB95} Hawley, J.~F., Gammie, 
C.~F., \& Balbus, S.~A.\ 1995, \apj, 440, 742 

\bibitem[Hawley et al.(2011)]{HGK} Hawley, J.~F., Guan, X., 
\& Krolik, J.~H.\ 2011, \apj, 738, 84

\bibitem[Heinemann 
\& Papaloizou(2009a)]{HP} Heinemann, T., \& Papaloizou, J.~C.~B.\ 2009, \mnras, 397, 52 

\bibitem[Heinemann 
\& Papaloizou(2009b)]{HP1} Heinemann, T., \& Papaloizou, J.~C.~B.\ 2009, \mnras, 397, 64 

\bibitem[Heinemann 
\& Papaloizou(2012)]{HP2} Heinemann, T., \& Papaloizou, J.~C.~B.\ 2012, \mnras, 419, 1085 

\bibitem[Inogamov 
\& Sunyaev(1999)]{InogamovSunyaev} Inogamov, N.~A., \& Sunyaev, R.~A.\
  1999, Astronomy Letters, 25, 269 

\bibitem[Koldoba et al.(2002)]{Koldobaetal} Koldoba, A.~V., 
Romanova, M.~M., Ustyugova, G.~V., \& Lovelace, R.~V.~E.\ 2002, \apjl,
576, L53 

\bibitem[Lai 
\& Tsang(2009)]{LaiTsang2009} Lai, D., \& Tsang, D.\ 2009, \mnras, 393, 979 

\bibitem[Landau 
\& Lifshitz(1959)]{LL} Landau, L.~D., \& Lifshitz, E.~M.\ 1959, Course
  of theoretical physics, Oxford: Pergamon Press, 1959

\bibitem[Larson(1990)]{Larson} Larson, R.~B.\ 1990,
  \mnras, 243 588

\bibitem[Lesur 
\& Longaretti(2005)]{LesurLongaretti} Lesur, G., \& Longaretti, P.-Y.\
  2005, \aap, 444, 25

\bibitem[Li et al.(2003)]{LGN} Li, L.-X., Goodman, J., 
\& Narayan, R.\ 2003, \apj, 593, 980

\bibitem[Li 
\& Narayan(2004)]{LN} Li, L.-X., \& Narayan, R.\ 2004, \apj, 601, 414  

\bibitem[Lynden-Bell 
\& Pringle(1974)]{LyndenBellPringle} Lynden-Bell, D., \& Pringle,
  J.~E.\ 1974, \mnras, 168, 603

\bibitem[Matsumoto 
\& Tajima(1995)]{MatsumotoTajima} Matsumoto, R., \& Tajima, T.\ 1995,
  \apj, 445, 767 

\bibitem[Miles(1958)]{MilesKH} Miles, J.~W.\ 1958, Journal of 
Fluid Mechanics, 4, 538

\bibitem[Narayan(1992)]{Narayan} Narayan, R.\ 1992, \apj, 394, 
261 

\bibitem[Narayan 
\& Popham(1993)]{NarayanPopham} Narayan, R., \& Popham, R.\ 1993,
  \nat, 362, 820

\bibitem[Ogilvie 
\& Pringle(1996)]{OgilviePringle} Ogilvie, G.~I., \& Pringle, J.~E.\
  1996, \mnras, 279, 152 

\bibitem[Pessah \& Chan(2012)]{Pessah} Pessah, M.~E., \& Chan, C.-k.\
  2012, \apj, 751, 48

\bibitem[Piro 
\& Bildsten(2004)]{PiroBildsten} Piro, A.~L., \& Bildsten, L.\ 2004,
  \apj, 610, 977 

\bibitem[Piro 
\& Bildsten(2007)]{PiroBildsten07} Piro, A.~L., \& Bildsten, L.\ 2007,
  \apj, 663, 1252 

\bibitem[Popham 
\& Narayan(1992)]{PophamNarayan1} Popham, R., \& Narayan, R.\ 1992, \apj, 394, 255 

\bibitem[Popham 
\& Narayan(1995)]{PophamNarayan} Popham, R., \& Narayan, R.\ 1995,
  \apj, 442, 337 

\bibitem[Pringle(1989)]{Pringle1989} Pringle, J.~E.\ 1989, \mnras, 
236, 107 

\bibitem[Rafikov(2012)]{Rafikov2012} Rafikov, R. R. 2012, arXiv:1205.5017 

\bibitem[Shakura 
\& Sunyaev(1973)]{ShakuraSunyaev} Shakura, N.~I., \& Sunyaev, R.~A.\
  1973, \aap, 24, 337

\bibitem[Sorathia et al.(2012)]{Sorathia} Sorathia, K.~A., 
Reynolds, C.~S., Stone, J.~M., \& Beckwith, K.\ 2012, \apj, 749, 189 

\bibitem[Spruit(1999)]{Spruit1999}	
Spruit, H. C. 1999, A\&A, 349, 189

\bibitem[Spruit(2002)]{Spruit2002}	
Spruit, H. C. 2002, A\&A, 381, 923

\bibitem[Steinacker 
\& Papaloizou(2002)]{SteinackerPapaloizou} Steinacker, A., \&
  Papaloizou, J.~C.~B.\ 2002, \apj, 571, 413

\bibitem[Stone et al(2008)]{Stone}	
Stone, J. M., Gardiner, T. A., Teuben, P., Hawley, J. F., \& Simon, J. B. 2008, ApJS, 178, 137

\bibitem[Tagger et al.(1992)]{TPC} Tagger, M., Pellat, R., 
\& Coroniti, F.~V.\ 1992, \apj, 393, 708 

\bibitem[Tayler(1973)]{Tayler} Tayler, R. J.\ 1992, \mnras, 161, 365 

\bibitem[Terquem \& Papaloizou(1996)]{TerquemPapaloizou} Terquem, C.,
  \& Papaloizou, J.~C.~B.\ 1996, \mnras, 279, 767

\bibitem[Tsang 
\& Lai(2009a)]{TsangLai0} Tsang, D., \& Lai, D.\ 2009, \mnras, 393, 992 

\bibitem[Tsang 
\& Lai(2009b)]{TsangLai} Tsang, D., \& Lai, D.\ 2009, \mnras, 396, 589

\bibitem[Tsang 
\& Lai(2009c)]{TsangLai1} Tsang, D., \& Lai, D.\ 2009,
  \mnras, 400, 470

\bibitem[Velikhov(1959)]{Velikhov} 
Velikhov, E. P. 1959 J. Exptl. Theoret. Phys., 36, 1398

\end{thebibliography}
\end{document}